\documentclass[%
 reprint,
 amsmath,amssymb,
 aps,
]{revtex4-2}

\usepackage{graphicx}%
\usepackage{dcolumn}%
\usepackage{multirow}
\usepackage{amssymb}
\usepackage{comment}
\usepackage{MnSymbol,wasysym}

\usepackage[dvipsnames,table,xcdraw]{xcolor}

\usepackage{hyperref}
\hypersetup{
    colorlinks=true,
    linkcolor=Blue,
    filecolor=Blue,      
    urlcolor=Blue,
    citecolor=blue,
}

\begin{document}

\title{
Superconductivity in disordered locally noncentrosymmetric materials: an application to \texorpdfstring{CeRh$_2$As$_2$}{}
}%

\author{David  M\"{o}ckli}

\affiliation{Instituto de F\'{i}sica, Universidade Federal do Rio Grande do Sul, 91501-970 Porto Alegre, Brazil}

\author{Aline Ramires}
\affiliation{Paul Scherrer Institut, CH-5232 Villigen PSI, Switzerland}

\date{\today}

\begin{abstract}
Layered three-dimensional centrosymmetric crystals can exhibit characteristics of noncentrosymmetric materials. This happens when each individual layer alone lacks inversion, but, when combined,  inversion symmetry is restored; hence the designation: \textit{locally noncentrosymmertic superconductors} (LNCSs). 
In LNCSs, the effects of impurities and subdominant magnetic field induced pairing channels remain unexplored. 
Using a minimal model, we examine all pairing channels and show that there is always a subdominant superconducting instability that is favored at high magnetic fields, which can substantially alter the magnetic field -- temperature phase diagram.
Also, we find that the phase diagram responds to disorder in a non-monotonic way, which can be subjected to experimental verification. 
We apply these ideas to the recently unveiled two-phase superconducting phase diagram of CeRh$_2$As$_2$.
We identify the two phases as singlet-triplet mixed even- and odd-parity states at low and a high fields, respectively.
Furthermore, we predict the presence of two superconducting phases also for in-plane magnetic fields in cleaner samples, since a high-field phase could have been so far hindered by impurity effects.
\end{abstract}

\maketitle


\section{Introduction}

Noncentrosymmetric superconductors have been extensively studied since the discovery of CePt$_3$Si \cite{Bauer:2004,Bauer:2012}.
In these materials, inversion symmetry breaking introduces an antisymmetric 
spin-orbit coupling (SOC) that lifts the spin degeneracy and can fundamentally affect the superconducting state. 
The non-trivial spin texture around the Fermi surface leads to the development of unusual properties, such as anisotropic spin susceptibility, enhanced Pauli limit,  spin singlet-triplet mixing, and magnetoelectric effect \cite{Frigeri:2004,Frigeri:2004b,Samokhin:2005,Edelstein:2005,Yip:2002,Fujimoto:2005,Fujimoto:2007}. Remarkably, the phenomenology of noncentrosymmetric superconductors can also be observed in centrosymmetric materials if these are formed by sub-units that locally break inversion symmetry  \cite{Maruyama:2012,Yoshida2012orig,Fischer:2011,yoshida2014,Sigrist2014}. 
In these systems, when SOC is comparable to, or larger than interlayer hopping (ILH) amplitudes, the effects of local noncentrosymmetricity manifest, and unconventional superconductivity can emerge at high magnetic fields \cite{Maruyama:2012,Yoshida2012orig,Yoshida2013complex}.  This possibility has been originally discussed in the context of multi-layer materials and heterostructures, such as artificial superlattices of CeCoIn$_5$ and YbCoIn$_5$ \cite{Shishido:2010, Mizukami:2011}.

\begin{figure}[t]
\includegraphics[width=0.8\linewidth, keepaspectratio]{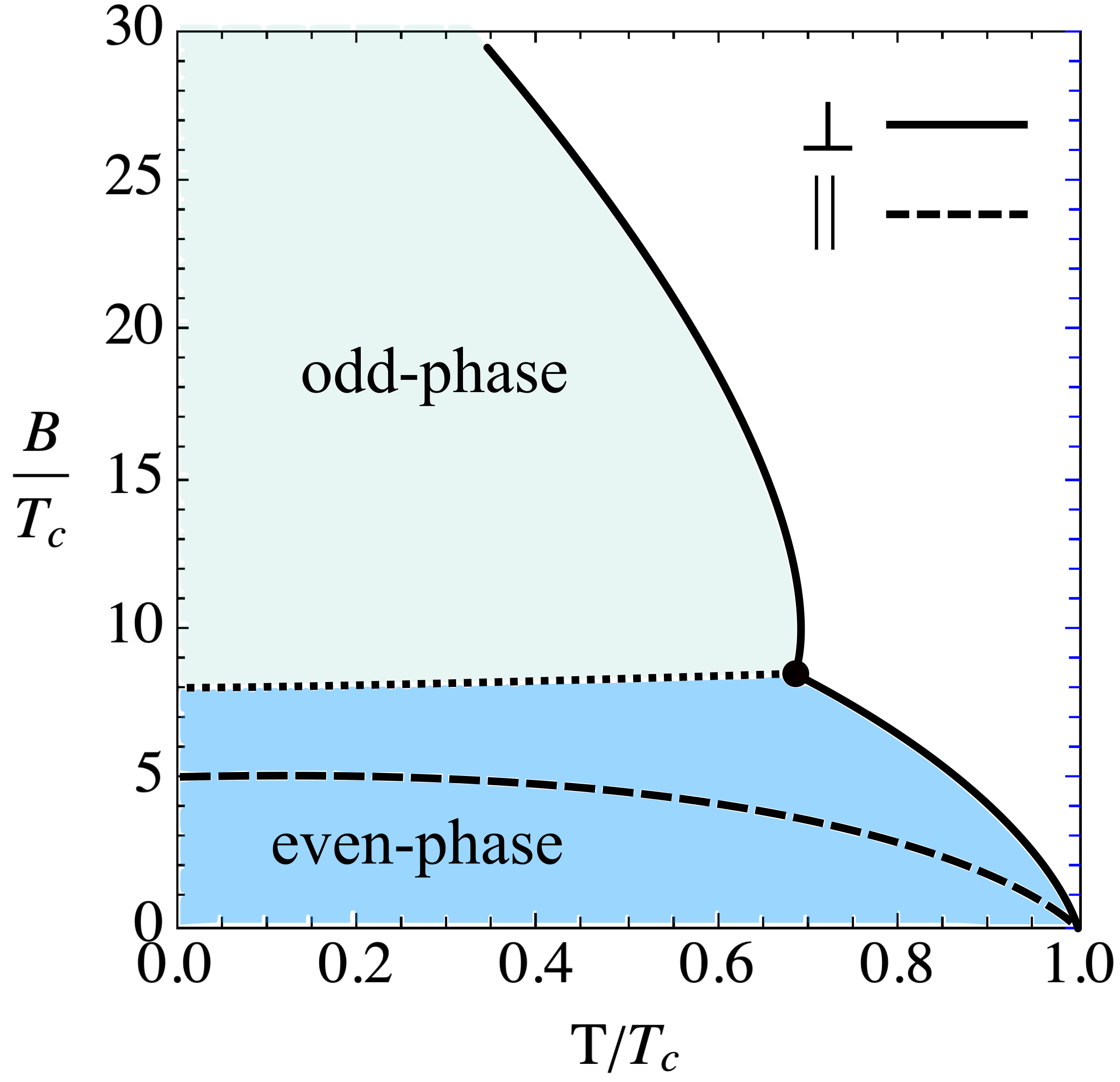}
\caption{\label{fig:diagram} 
Schematic $(B,T)$ phase diagram of CeRh$_2$As$_2$. The dashed line shows the transition line for in-plane ($\parallel$) magnetic field. The solid line indicates the transition line for the $c$-axis ($\perp$) magnetic field. If the low and high-field phases have different superconducting order parameters, the dot indicates a multicritical point that requires a first-order phase transition separating the two phases (dotted line). 
All quantities have units of energy. To recover the international units, make $B/T_c\rightarrow \mu_\mathsf{B}B/(k_\mathsf{B}T_c)$.
}
\end{figure}

Recent experiments on the locally noncentrosymmetric superconductor (LNCS) and heavy fermion CeRh$_2$As$_2$ unveil a rare magnetic field versus temperature phase diagram with two superconducting phases \cite{khim2021fieldinduced} (see Fig. \ref{fig:diagram}). 
Under a $c$-axis magnetic field, a phase transition from a low-field to a high-field superconducting phase occurs around 4T, and the upper critical field in this direction reaches up to 14T, much above the Pauli limit $H_\mathsf{P} \approx$ 0.5T for a superconductor with a critical temperature ($T_c$) of 0.26K.
This type of phase diagram has been predicted by theories developed for LNCS superconductors \cite{Fischer:2011,Maruyama:2012,Yoshida2012orig,Sigrist2014,yoshida2014,Higashi2016,nakamura2017}, but was not observed in any previous material or heterostructure lacking inversion symmetry locally. 
The observation of this unique two-phase phase diagram in CeRh$_2$As$_2$ generates multiple questions, and has caused a revival of research in LNCS  \cite{nogaki2021topological,skurativska2021spin,ptok2021electronic,Cavanagh:2021,schertenleib2021unusual}.

In the standard theories for LNCS, conventional spin singlet pairing is assumed to be the stable superconducting state within each layer. Once a magnetic field is applied perpendicularly to the layers, a pair-density wave (PDW) state, with a sign change of the order parameter between layers is favored under the requirement that Rashba SOC is larger than ILH amplitudes \cite{Yoshida2012orig,Yoshida2013complex,yoshida2014,Sigrist2014,nakamura2017}. 
However, CeRh$_2$As$_2$ displays only a small effective mass anisotropy (inferred from the slope of the upper critical field around the critical temperature \cite{khim2021fieldinduced}), indicating that the system is rather three-dimensional. 
This is corroborated by recent first-principles calculations \cite{ptok2021electronic,nogaki2021topological}, and by comparison to other 122-materials in this family. As an example, CeCu$_2$Si$_2$ crystallizes in the ThCr$_2$Si$_2$-type structure, the locally centrosymmetric analog of the CaBa$_2$Ge$_2$-type structure of CeRh$_2$As$_2$, and displays a three-dimensional spin density wave state supported by Fermi surface nesting with superconductivity emerging around the pressure induced quantum critical point \cite{Arndt:2011,Steglich:2012}. This leaves us with the question: are there other key ingredients besides local inversion symmetry breaking for the development of the unusual phase diagram displayed by CeRh$_2$As$_2$?

Heavy fermions usually exhibit rich phase diagrams displaying heavy Fermi liquid behavior, magnetic and superconducting phases, as well as ``hidden orders", associated with the development of multipolar order parameters \cite{Paschen:2021}.
The small energy scales associated with the ordered states make these systems highly tunable and prone to quantum critical phenomena \cite{RamiresNP:2019,Si:2010}. In this context, the specific heat coefficient of CeRh$_2$As$_2$ follows an unusual power-law dependence in temperature below 4K \cite{khim2021fieldinduced}, indicating the proximity to a quantum critical point. The presence of quantum fluctuations is usually connected with the development of unconventional superconductivity in these materials \cite{Dyke:2014,Landaeta:2018, White:2015,Smidman:2018}. These facts raise the question: is pairing in unconventional channels important for the development of such two-phase phase diagram?

Motivated by these questions, here we build up on our previous work \cite{mockli2021scenarios} and revisit the problem of layered superconductors from a microscopic perspective. Our goal is to explore the effects of normal state parameters, magnetic fields, the presence of sub-leading superconducting instabilities, and impurities on the phase diagram of LNCS, with focus on CeRh$_2$As$_2$. 
We highlight the competition between inter-sublattice hopping (ILH) and spin-orbit coupling (SOC), explore the effects of channel mixing promoted by magnetic field, and investigate the effects of non-magnetic impurities in the low and high magnetic field phases. 
For this, we employ the linearized quasiclassical Eilenberger formalism, which allows us to gain analytical insights and perform a guided exploration of the parameter space.

This work is organized as follows. In Sec. \ref{sec:model} we introduce the minimal Bogoliubov-deGennes Hamiltonian, written in terms of the most general normal state model based on a sublattice degree of freedom (DOF) associated with the LNCS structure. 
We also introduce all possible superconducting states and highlight their properties. 
In Sec. \ref{sec:quasiclassical}, we develop the corresponding quasiclassical theory,  including the effect of isotropic scalar impurities. 
We derive the linearized Eilenberger equations and solve them analytically for an appropriate set of parameters. The solutions of the Eilenberger equations provide us with the superconducting -- normal state transition lines. 
In Sec. \ref{sec:phasediagrams} we obtain the complete phase diagram and discuss the role of the sub-leading triplet channels.
In Sec. \ref{sec:impurities} we study how nonmagnetic impurities affect the phase diagram. 
The discussion in Sec. \ref{sec:discussion} examines the parameters that are consistent with experiment and identifies aspects that require further investigation.
In addition, we highlight potentially interesting magnetic field effects in these systems that could be the topic of future work. 
Appendices \ref{app:low} and \ref{app:high} provide more detailed derivations of the results in the main text.

\section{The model\label{sec:model}}

We start with the minimal Bogoliubov-deGennes (BdG) mean-field framework able to capture a LNCS structure by including a sublattice DOF, in addition to the spin DOF. The BdG Hamiltonian can then be expressed as an $8 \times 8$ matrix
\begin{align}
\mathcal{H}_\mathsf{BdG} (\mathbf{k}) = \boldsymbol{\Psi}^\dag_\mathbf{k} 
\begin{bmatrix}
\hat{H}_0(\mathbf{k}) & \hat{\Delta}(\mathbf{k})\\ 
\hat{\Delta}^\dag(\mathbf{k}) & -\hat{H}_0^*(-\mathbf{k})
\end{bmatrix}
\boldsymbol{\Psi}_\mathbf{k},
\label{eq:hbdg}
\end{align}
where $\boldsymbol{\Psi}^\dag_\mathbf{k} = (\boldsymbol{\Phi}_\mathbf{k}^\dag,\boldsymbol{\Phi}^\mathsf{T}_{-\mathbf{k}})
$ is a Nambu vector with $\boldsymbol{\Phi}^\dag_\mathbf{k} = (c^\dag_{1\mathbf{k}\uparrow},c^\dag_{1\mathbf{k}\downarrow},c^\dag_{2\mathbf{k}\uparrow},c^\dag_{2\mathbf{k}\downarrow} )
$. Here, the operator $c^\dagger_{n\mathbf{k}s}$ ($c_{n\mathbf{k}s}$) corresponds to the creation (annihilation) of an electron at the $n$ sublattice with momentum $\mathbf{k}$ and $z$-spin projection $s=\uparrow,\downarrow$. $\hat{H}_0(\mathbf{k})$ is the normal state Hamiltonian and $\hat{\Delta}(\mathbf{k})$ the gap matrix.

Throughout the remainder of the text, we set the Boltzmann constant and Bohr magneton $k_\mathsf{B}=\mu_\mathsf{B}=1$, such that all quantities have units of energy, and we absorb the $g$-factor into the magnetic induction $\mathbf{B}$. 

\begin{figure}
\includegraphics[width=0.98\linewidth, keepaspectratio]{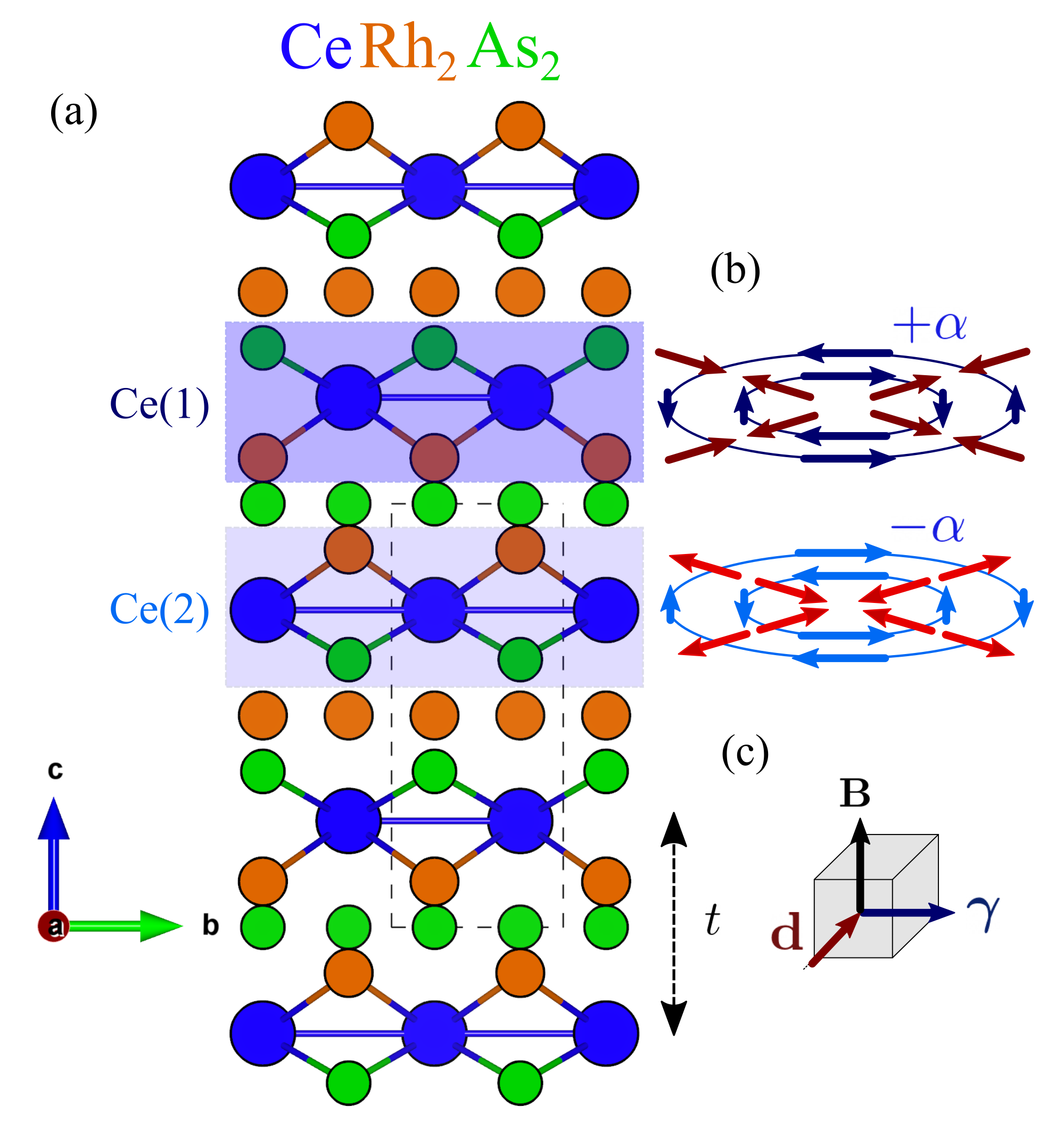}
\caption{\label{fig1:crystal} 
(a) Crystal structure of CeRh$_2$As$_2$. 
The dashed box indicates the centrosymmetric unit cell. 
The sticks connect the Ce atoms to the nearest Rh and As, which shows the different Ce(1) and Ce(2) sublattice environments. The shaded boxes indicate the two inversion-broken local environments.
(b) Schematic Fermi surfaces for the independent Ce sublattices. The blue arrows are the Rashba spin-texture, which have opposite directions in Ce(1) and Ce(2). 
The red arrows show the texture of the magnetic field induced spin-triplet $d$-vector.
(c) The relative orientation of magnetic field, Rashba SOC, and field induced triplets is indicated in the parallelepiped, whose volume $\mathbf{B}\times\boldsymbol{\gamma}(\mathbf{k})\cdot \mathsf{Im}\,\mathbf{d}(\mathbf{k})$  gives the singlet-triplet coupling. 
The figure was produced with the aid of VESTA \cite{Momma2011}.
}
\end{figure}

\subsection{The normal state Hamiltonian \label{sec:normal_state}}

The normal state Hamiltonian is a $4\times 4$ matrix in sublattice $\otimes$ spin space. Focusing on CeRh$_2$As$_2$, we can construct a specific normal state Hamiltonian considering the details of its structure, characterized by the space group P4/nmm (No.  129). While the crystal is centrosymmetric, the atomic positions lack inversion symmetry.

Given the heavy fermion nature of the electronic structure,
we model the electronic degrees of freedom from the Ce sites perspective. There are two inequivalent types of layers of Ce sites: for the first type, Ce(1) sites are coordinated with As on top and Rh at the bottom, while for the second type, Ce(2), Rh and As atoms are exchanged, as shown in Fig. \ref{fig1:crystal}. 
The two inequivalent Ce sites give origin to a sublattice structure as an internal degree of freedom. The two inequivalent  Ce sites  have a reduced $C_{4v}$ point group  symmetry and are not centers of inversion. The center of inversion lies at the midpoint between the two inequivalent Ce sites. Therefore, inversion exchanges the sublattices.

The most general normal state Hamiltonian for CeRh$_2$As$_2$ can be written as (here we omit the Kronecker product $\otimes$)
\begin{align}
\hat{H}_0(\mathbf{k})=\xi(\mathbf{k})\tau_0\sigma_0+\mathbf{t}(\mathbf{k})\cdot\boldsymbol{\tau}\sigma_0+\left[\tau_3\boldsymbol{\gamma}(\mathbf{k})-\tau_0\mathbf{B} \right ]\cdot\boldsymbol{\sigma},
\end{align}
Here, $\boldsymbol{\tau}=(\tau_1,\tau_2,\tau_3)$ and $\boldsymbol{\sigma}=(\sigma_1,\sigma_2,\sigma_3)$ are the Pauli matrices vectors in sublattice and spin-space, respectively.  Following Refs. \cite{khim2021fieldinduced,mockli2021scenarios}, $\xi(\mathbf{k})$ describes intra-sublattice hopping processes. The vector $\mathbf{t}(\mathbf{k})=(t_1(\mathbf{k}),t_2(\mathbf{k}),0)$ describes the symmetry allowed ILH processes:
\begin{align}
    t_1(\mathbf{k})& =c_1t_1\cos\left(\frac{k_x a}{2}\right)\cos\left(\frac{k_y a}{2}\right)\cos\left(\frac{k_z c}{2}\right); \notag \\
    t_2(\mathbf{k}) & = c_2t_2\cos\left(\frac{k_x a}{2}\right)\cos\left(\frac{k_y a}{2}\right)\sin\left(\frac{k_z c}{2}\right),
\end{align}
where $c_1$ and $c_2$ are normalization constants such that $\langle t^2_{1(2)}(\mathbf{k})\rangle_\mathbf{k}=t^2_{1(2)}$, where $\langle \ldots\rangle_\mathbf{k}$ indicates the average over the Fermi surface. Note that $t_2(\mathbf{k})$ is present due to the local inversion symmetry breaking. The vector $\boldsymbol{\gamma}(\mathbf{k})$ is odd in momentum and corresponds to a staggered intra-sublattice SOC that also arises due to local inversion symmetry breaking. The accompanying $\tau_3$ matrix ensures global inversion symmetry. The crystal structure imposes: $\boldsymbol{\gamma}(\mathbf{k})= (\gamma_x(\mathbf{k}),\gamma_y(\mathbf{k}),\gamma_z(\mathbf{k}))$, where $\left\{\gamma_x(\mathbf{k}),\gamma_y(\mathbf{k}) \right\}= c_\alpha\alpha\left\{\sin(k_y a),-\sin(k_x a)\right\}$ and $\gamma_z(\mathbf{k})= c_\lambda\lambda \sin(k_x a)\sin(k_y a)\sin(k_z a)\left[\cos(k_x a)-\cos(k_y a)\right]$. The amplitudes $\alpha$ and $\lambda$ refer to the Rashba and Ising components and $c_{\alpha(\lambda)}$ are the respective normalization constants. Because $\lambda$ arises due to next nearest sublattice processes,  we expect $\alpha>\lambda$. Finally, $\mathbf{B}$ is the Zeeman magnetic field. 

\subsection{The superconducting order parameters}

\begin{figure}
\includegraphics[width=\linewidth, keepaspectratio]{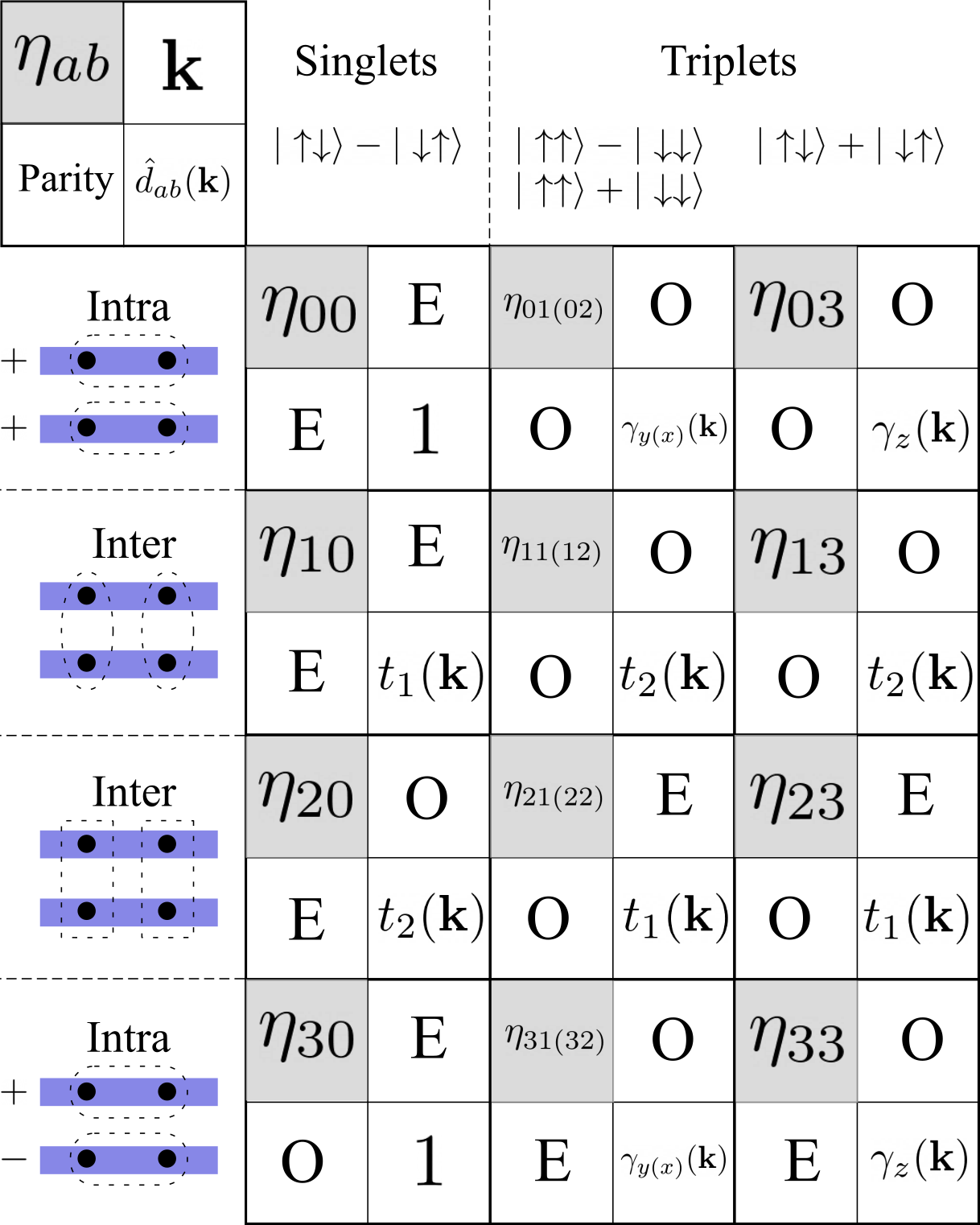}
\caption{\label{fig:tabela}
Properties of all superconducting order parameters $\eta_{ab}$. The sublattice index $a$ labels the rows, and the spin index $b$ the columns. The information of an order parameter is organized in $2\times 2$ blocks that contain the order parameter $\eta_{ab}$, whether it is even (E) or odd (O) in $\mathbf{k}$ and parity, and the basis function $\hat{d}_{ab}(\mathbf{k})$. 
Given the sublattice exchange under inversion symmetry, the parity operator is implemented as $P=\tau_1\otimes \sigma_0$, accompanied by inversion of the momenta in $d_{ab}(\mathbf{k})$. 
The cartoons in the first column illustrate the different character of the order parameters with intra- or inter-sublattice Cooper pairing.
The column with $b=\{1,2\}$ lists the equal-spin triplets that are induced by a $c$-axis magnetic field. 
}
\end{figure}

From the microscopic perspective in sublattice $\otimes$ spin space, any superconducting order parameter can be cast as a matrix of the form
\begin{align}
\label{eq:Delta}
\hat{\Delta}(\mathbf{k})=\sum_{a,b=0}^3 \eta_{ab}\hat{d}_{ab}(\mathbf{k})\tau_a\otimes\sigma_b\,i\sigma_2.
\end{align}
Here, $\eta_{ab}$ carries the magnitude and phase of the order parameter and $\hat{d}_{ab}(\mathbf{k})$ is a normalized function of momentum. The order parameters $\eta_{ab}$ with $b=0$ correspond to the spin singlet Cooper pairs, whereas the order parameters with $b=1,2,3$ parametrize the spin triplets. Fermionic exchange requires $\hat{\Delta}(\mathbf{k})=-\hat{\Delta}^\mathsf{T}(\mathbf{-k})$, such that an order parameter with an antisymmetric (symmetric) matrix structure $\tau_a\otimes\sigma_b\,i\sigma_2$ is necessarily accompanied by a function $\hat{d}_{ab}(\mathbf{k})$  that is even (odd) in momentum. Note that, due to the extra sublattice degree of freedom, besides the usual momentum-even spin-singlet and momentum-odd spin-triplet order parameters, a momentum-odd spin-singlet or a momentum-even spin-triplet order parameter is allowed if it is antisymmetric in the sublattice degree of freedom. Fig. \ref{fig:tabela} summarizes all possible order parameters within this model and their properties.

\section{Quasiclassical theory \label{sec:quasiclassical}}

The Gor'kov Green's function $\mathsf{G}(\mathbf{k};\omega_n)$ corresponding to the Hamiltonian in Eq. \eqref{eq:hbdg} satisfies the Gor'kov equation
\begin{align}
    \left[i\omega_n\hat{\mathsf{1}}-\mathcal{H}_\mathsf{BdG}(\mathbf{k}) \right ]\mathsf{G}(\mathbf{k};\omega_n)=\hat{\mathsf{1}},
    \label{eq:gorkov}
\end{align}
where $\omega_n=(2n+1)\pi T$ ($n\in\mathbb{Z}$) are the fermionic Matsubara frequencies. We define the identity matrix  $\hat{\mathsf{1}}=\rho_0\otimes \tau_0\otimes\sigma_0$, where $\rho_0$ is the $2\times 2$ identity in Nambu  (particle-hole) space. We can supplement it with $\boldsymbol{\rho}=(\rho_1,\rho_2,\rho_3)$ to describe the complete Nambu space. 

\subsection{The Eilenberger matrix equation}

To develop the quasiclassical Eilenberger equations corresponding to the BdG Hamiltonian in Eq. \eqref{eq:hbdg}, we should identify which energy scales are expected to be small compared to the Fermi energy $E_\mathsf{F}$, which is of the order ($\sim$) of eV's \cite{ptok2021electronic}. 
For a magnetic field along the $c$-axis, the upper critical field in CeRh$_2$As$_2$ is $\sim \mu_\mathsf{B} B \approx 6\times 10^{-4}$eV. It is reasonable to assume that all spin related energy scales $B,\alpha,\lambda\ll E_\mathsf{F}$. Also, from the crystal structure, see Fig. \ref{fig1:crystal}, it is reasonable to assume that the energy scales of intra-sublattice processes are larger than ILH processes \cite{mockli2021scenarios}. 
To treat ILH and SOC on equal footing, we also consider the ILH energy scales  $t_1,t_2\ll E_\mathsf{F}$. Within a weak coupling perspective, we assume that the superconducting order parameters energy scales $T_c\ll E_\mathsf{F}$.
With the hierarchy of energy scales set, the resulting quasiclassical theory elucidates the interplay of SOC, magnetic field, ILH and superconductivity. 
We will see that the relevant phenomenology is realized for 
\begin{align}
    T_c < t_1,t_2 < \lambda,\alpha \ll E_\mathsf{F}.
    \label{eq:regime}
\end{align}

To obtain the Eilenberger equation,
we manipulate the Gor'kov equation Eq. \eqref{eq:gorkov} following analogous steps as detailed in Ref. \cite{Mockli2020}. The procedure yields
\begin{align}
\left[\left(i\omega_n\hat{\mathsf{1}}-\mathcal{H}(\mathbf{k}) \right )\rho_3,\rho_3\mathsf{G}(\mathbf{k};\omega_n) \right ]=0,
\label{eq:grokov_eilenberger}
\end{align}
where $\xi(\mathbf{k})$ is now removed from the problem by $\mathcal{H}(\mathbf{k})=\mathcal{H}_\mathsf{BdG}(\mathbf{k})-\xi(\mathbf{k})\rho_3\tau_0\sigma_0$.  
We introduce the dimensionless quasiclassical Green's functions \cite{Kopnin2001,Kita2015}
\begin{align}
g(\mathbf{k}_\mathsf{F};\omega_n) & =\oint \frac{\mathrm{d}\xi_\mathbf{k}}{\pi}\,i\rho_3\,\mathsf{G}(\mathbf{k};\omega_n)  \notag \\
& =
\begin{bmatrix}
\hat{g}(\mathbf{k}_\mathsf{F};\omega_n) & -i\hat{f}(\mathbf{k}_\mathsf{F};\omega_n)\\ 
-i\hat{f}^*(-\mathbf{k}_\mathsf{F};\omega_n) & -\hat{g}^*(-\mathbf{k}_\mathsf{F};\omega_n)
\end{bmatrix}.
\end{align}
Unlike the Gor'kov Green's function $\mathsf{G}(\mathbf{k};\omega_n) $, the quasiclassical Green's function is evaluated only at the Fermi momentum $\mathbf{k}_\mathsf{F}$. We henceforth drop the $\mathsf{F}$ subscript. 

Within the self-consistent Born approximation \cite{Kita2015,Mckli2020magnetic}, we also can include the effects of momentum-isotropic scalar impurities via the prescription $\mathcal{H}^0_\mathsf{BdG}(\mathbf{k}) \rightarrow \mathcal{H}^0_\mathsf{BdG}(\mathbf{k})+\Sigma (\omega_n)$, where the impurity self-energy is given by
\begin{align}
\Sigma (\omega_n) = -i\Gamma\left\langle g(\mathbf{k};\omega_n) \right\rangle_\mathbf{k}\rho_3,
\label{eq:impurity_selfenergy}
\end{align}
where $\Gamma$ is the scalar impurity scattering rate.
With these elements, we can now write the Eilenberger matrix equation from Eq. \eqref{eq:grokov_eilenberger} as
\begin{align}
\left[\left(i\omega_n\hat{\mathsf{1}}-\mathcal{H}(\mathbf{k})-\Sigma(\omega_n) \right )\rho_3,g(\mathbf{k};\omega_n) \right ]=0.
\label{eq:eilenberger_matrix}
\end{align}
Together with the normalization condition $g^2(\mathbf{k};\omega_n)=\hat{\mathsf{1}}$, Eq. \eqref{eq:eilenberger_matrix} determines all elements of $g(\mathbf{k};\omega_n) $.

\subsection{ The linearized Eilenberger matrix equation}

We are ultimately interested in the superconducting transition lines in the $(B,T)$ phase diagram. For this, we linearize the Eilenberger Eq. \eqref{eq:eilenberger_matrix} and solve for the superconducting correlations $\hat{f}(\mathbf{k};\omega_n)$. 
We parametrize the correlations by
\begin{align}
    \hat{g}(\mathbf{k};\omega_n) & = \sum_{a,b=0}^3 g_{ab}(\mathbf{k};\omega_n)\tau_a\otimes \sigma_b; \notag \\
   \hat{f}(\mathbf{k};\omega_n) & = \sum_{a,b=0}^3 f_{ab}(\mathbf{k};\omega_n)\tau_a\otimes \sigma_b\,i\sigma_2 .
\end{align}
Linearization together with the normalization condition imposes that only $g_{00}(\mathbf{k};\omega_n)=g^*_{00}(-\mathbf{k};\omega_n)=\mathsf{sgn}(\omega_n)$ in $\hat{g}$ survives. 
Here, $\mathsf{sgn}(\omega_n)$ is the sign function.
With this, we write the linearized Eilenberger equation
\begin{align}
 2i\bar{\omega}_n\hat{f}-\hat{H}(\mathbf{k})\hat{f}+\hat{f}\hat{H}^*(-\mathbf{k})=2i\mathsf{sgn}(\omega_n)\bar{\Delta}(\mathbf{k}), \label{eq:linEilenberger}
\end{align}
where
\begin{align}
& \bar{\omega}_n=\omega_n+\Gamma\mathsf{sgn}(\omega_n);  \\
& \bar{\Delta}(\mathbf{k}) = \hat{\Delta}(\mathbf{k}) +\Gamma \langle \hat{f}(\mathbf{k};\omega_n)\rangle_\mathbf{k}.
\end{align}
Eq. \eqref{eq:linEilenberger} is a $4\times 4$ matrix equation, which we can solve for all sixteen $f_{ab}$'s in terms of their averages.
If one solves the problem in the clean case, the substitutions $\omega_n\rightarrow \bar{\omega}_n$ and $\hat{\Delta}(\mathbf{k})\rightarrow\bar{\Delta}(\mathbf{k})$ yield the solutions in terms of the impurity averages. 

\subsection{The self-consistency and pair-breaking condition}

Finally, the solutions for the correlations $f_{ab}(\mathbf{k};\omega_n)$ are supplied to the self-consistency condition for the order parameters
\begin{align}
\eta_{ab}\ln\frac{T}{T_\mathrm{ab}}+\pi T\sum_{n\in\mathbb{Z}}
\left[\frac{\eta_{ab}}{|\omega_n|}  -\left\langle\hat{d}_{ab}(\mathbf{k})f_{ab}(\mathbf{k};\omega_n)\right\rangle_{\mathbf{k}} \right] =0.
\label{eq:self_consistency} 
\end{align}
Here, in the most general case, each order parameter $d_{ab}(\mathbf{k})=\eta_{ab}\hat{d}_{ab}(\mathbf{k})$ has an associated superconducting critical temperature $T_{ab}$, which is defined in the absence of SOC, magnetic field and ILH processes. 
The $T_{ab}$'s are defined in favor of the dimensionless coupling strength $\lambda_{ab}=\ln\left[2e^\gamma\epsilon_\mathsf{c}/(\pi T_{ab})\right]$, where  $\epsilon_\mathsf{c}$ is a characteristic cutoff energy of the pairing interaction \cite{Kita2015,mockli2019,Mockli2020}. 

It is interesting to note that the modified commutator, $\hat{H}(\mathbf{k})\hat{f}-\hat{f}\hat{H}^*(-\mathbf{k})$, in Eq. \eqref{eq:linEilenberger} 
relates to pair-breaking effects, and resembles the superconducting fitness measure \cite{ramires2016,ramires2018}.
If this modified commutator vanishes, then $\langle\hat{d}_{ab} f_{ab}\rangle_\mathbf{k}=\eta_{ab}/|\omega_n|$, which when substituted into Eq. \eqref{eq:self_consistency} yields $T=T_{ab}$. This means that there are no pair-breaking effects and that the order parameter is completely compatible with the underlying electronic structure.

\section{Phase diagrams \label{sec:phasediagrams}
}

Among the sixteen order parameter possibilities listed in Fig. \ref{fig:tabela}, the superconducting fitness analysis for the regime specified in Eq. \eqref{eq:regime} suggests four dominant order parameter candidates: $\{\eta_{00},\eta_{13},\eta_{30},\eta_{23}\}$ \cite{mockli2021scenarios}. 
However, because the triplet candidates $\{\eta_{13},\eta_{23}\}$ have a $d$-vector that is perpendicular to the layers, they are not limited for in-plane magnetic fields, which is inconsistent with the experimental data \cite{khim2021fieldinduced,skurativska2021spin}. 
Therefore, even tough we have no knowledge so far on the pairing mechanism, we are left with only $\{\eta_{00},\eta_{30}\}$ as suitable dominant candidates, which we examine in detail in the next subsection. 

We also investigated the regime $t_1,t_2>\alpha,\lambda$ for which the effects related to local noncentrosymmetricity loose relevance. In this case, the fitness analysis suggests a dominance of the $\eta_{01(02)}$ and $\eta_{11(12)}$ triplets. 
However, these triplets leave the critical field enhancement for an in-plane direction unexplained.

As we shall see, $\eta_{00}$ is the primary order parameter describing the low-field phase of CeRh$_2$As$_2$. According to the properties listed in Fig. \ref{fig:tabela},
the $\eta_{00}$ order parameter is: intra-sublattice, spin-singlet, momentum-even and parity-even. We henceforth simply refer to $\eta_{00}$ dominated states as the \textit{even-phase}. Similarly, $\eta_{30}$ is the primary order parameter describing the high-field phase and is an intra-sublattice, spin-singlet, momentum-even, and parity-odd order parameter. It is also called a pair-density wave (PDW) because of the staggered phase of the order parameter across the layers. We simply refer to $\eta_{30}$ dominated states as the \textit{odd-phase}.

\begin{figure*}
\includegraphics[width=\linewidth, keepaspectratio]{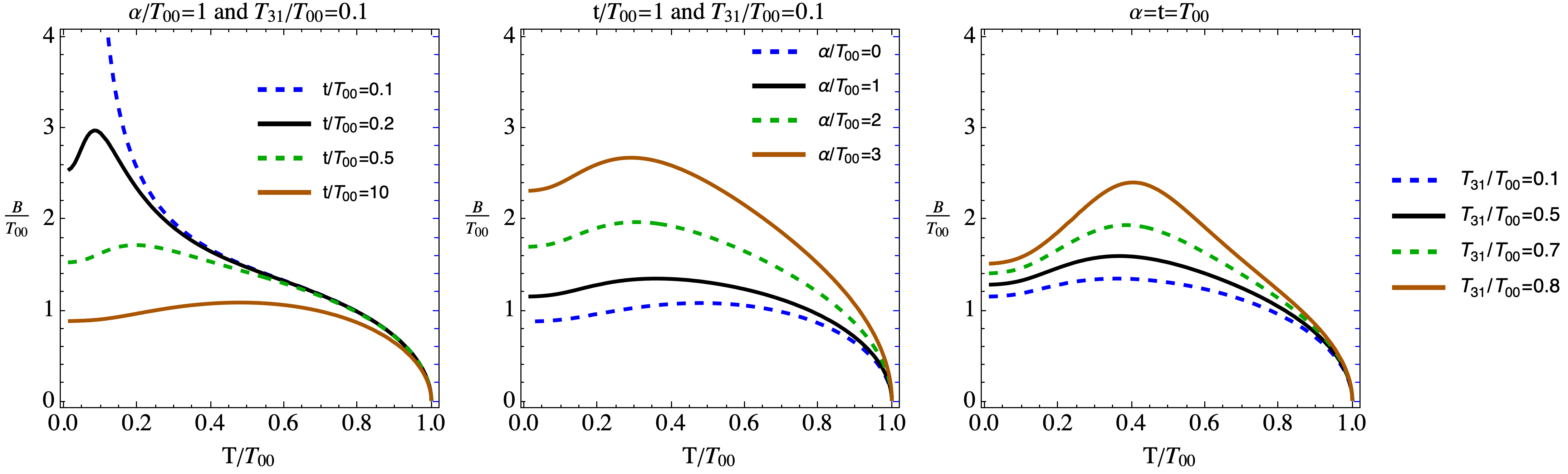}
\caption{\label{fig3:low}
Transition lines of the low-field even-phase $\{\eta_{00},\eta_{31(32)}\}$.
The heading titles of each plot indicate which parameters are fixed, and the line-legends show the parameter that changes.
(a) Suppression of the critical field for larger ILH $t$.
(b) Enhancement of the critical field for larger SOC.
(c) Enhancement of the upper critical field by the presence of subleading instabilities for the magnetic field induced triplets with critical temperature $T_{31(32)}$. The enhancement due to the subleading triplets is only significant for $t\lesssim T_{00}$.
}
\end{figure*}

\subsection{ The low field even-phase
\label{sec:low}}

To illustrate the general procedure, we solve the linearized Eilenberger matrix equation \eqref{eq:linEilenberger} for $\mathbf{B}=(0,0,B)$ in more detail here than in subsequent sections. For a magnetic field along the $c$-axis, it is the Rashba SOC component $\alpha$ that enhances the critical field.
For this reason, we set the Ising component $\lambda=0$ in this section. Although we need to solve a system of sixteen equations given by Eq. \eqref{eq:linEilenberger}, analytic solutions are possible for $t_1=0$ or $t_2=0$. The conclusions of this paper do not depend on whether $t_1$, $t_2$ or both are included. However, the calculations simplify significantly if only one of them is treated in the analysis. For this reason, we choose to treat $t_1$ and $t_2$ separately. Solving the system for all $f_{ab}$'s, here we write the solution for $f_{00}$. We omit the momentum and frequency arguments for conciseness, and obtain
\begin{widetext}
\begin{align}
 f_{00} = \frac{|\bar{\omega}_n|\left(\bar{\omega}_n^2+\boldsymbol{\gamma}^2+t_{1(2)}^2 \right )\bar{d}_{00}+i\mathsf{sgn}(\omega_n)B\left[ \left(\bar{\omega}_n^2+t_{1(2)}^2 \right )\bar{d}_{03}-\gamma_y\left(\pm t_{1(2)}\bar{d}_{21(11)}+\bar{\omega}_n \bar{d}_{31} \right )+\gamma_x\left(\pm t_{1(2)}\bar{d}_{22(12)}+\bar{\omega}_n \bar{d}_{32} \right )\right ]}{\left(\bar{\omega}_n^2+t_{1(2)}^2 \right )\left(\bar{\omega}_n^2+B^2 \right )+\bar{\omega}_n^2\boldsymbol{\gamma}^2}.  
\label{eq:f00}
\end{align}
\end{widetext}
Here, $\bar{d}_{ab}(\mathbf{k})=d_{ab}(\mathbf{k})+\Gamma \langle f_{ab}(\mathbf{k};\omega_n)\rangle_\mathbf{k}$. 
In the terms with $\pm t_{1(2)}$, $+t_1$ corresponds to the case when only $t_1$ is present, and $-t_2$ corresponds to the case when only $t_2$ is included. 
Note that even if we only allow the $d_{00}$ order parameter to exist (by setting all other $d_{ab}=0$), some $\bar{d}_{ab}$ channels might still acquire a finite value due to impurities. In the following, we focus on the clean solutions of $\eta_{00}$ first, and make no assumption on the vanishing of other order parameters that might couple to the dominant $\eta_{00}$.

To obtain the transition line for the even-phase in the $(B,T)$ phase diagram in the clean case ($\Gamma=0$), we should feed Eq. \eqref{eq:f00} to the self-consistency condition in Eq. \eqref{eq:self_consistency}. 
If we choose an $s$-wave-like state for $\hat{d}_{00}=1$, such that $d_{00}=\eta_{00}\hat{d}_{00}=\eta_{00}$, the average of Eq. \eqref{eq:f00} yields
\begin{align}
\langle f_{00} \rangle_\mathbf{k}=|\omega_n|
\frac{\left(\omega_n^2+\alpha^2+t^2 \right )\eta_{00}+i\sqrt{2}B\alpha\eta_{31(32)}}{(\omega_n^2+t^2)(\omega_n^2+B^2)+\omega_n^2\alpha^2};
\label{eq:av_f00}
\end{align}
When we take the average leading to Eq. \eqref{eq:av_f00}, instead of performing the angular integrations exactly, 
a good approximation is to perform the substitutions $\langle \boldsymbol{\gamma}^2(\mathbf{k})d_{00}(\mathbf{k}) \rangle_\mathbf{k}\rightarrow\alpha^2 \langle d_{00}(\mathbf{k})\rangle_\mathbf{k}$ and 
$\langle t_{1(2)}^2(\mathbf{k})\rangle_\mathbf{k}\rightarrow t^2$.
These approximations are discussed in detail in Ref. \cite{Mockli2020}, and allow us to make analytical progress, yet still obtain qualitatively correct results.
Also, we used the same basis function harmonic for the triplet components as the Rashba-SOC (see Appendix \ref{app:low} for more details)
\begin{align}
d_{31(32)}(\mathbf{k}) & =\eta_{31(32)}\hat{d}_{31(32)}(\mathbf{k}); \\
\hat{d}_{31(32)}(\mathbf{k}) &
=\mp  \sqrt{2}\hat{\gamma}_{y(x)}(\mathbf{k}),
\label{eq:d31}
\end{align}
In fact, this is the only basis function harmonic that couples to the singlets; see red $d$-vector textures in Fig. \ref{fig1:crystal}. 
The joint action of SOC and magnetic field selects the harmonic in Eq. \eqref{eq:d31}.
One can think of the magnetic field converting the singlets that initially have the same texture as SOC into equal-spin triplets.
The converted triplets contribute to the triple product $\mathbf{B}\times\boldsymbol{\gamma}(\mathbf{k})\cdot \mathsf{Im}\,\mathbf{d}(\mathbf{k})$; according to the  parallelepiped in Fig. \ref{fig1:crystal}(c). 
In Fig. \ref{fig:tabela}, we summarize the symmetry properties of the superconducting order parameters. These properties are extensively used to obtain the relevant correlation averages that enter the self-consistency condition \eqref{eq:self_consistency}.

Before obtaining the $(B,T)$ transition curve, 
let us check the known limits of Eq. \eqref{eq:av_f00}. 
If $B=0$, then $\langle f_{00}\rangle =\eta_{00}/|\omega_n|$ such that triplets, SOC and ILH have no bearing on the $\eta_{00}$ singlets. 
If $\alpha=0$, then $\langle f_{00}\rangle =|\omega_n|\eta_{00}/(\omega_n^2+B^2)$, which leads to Pauli limiting of the singlets. 
If $t=0$, the sublattices decouple and the physics maps to the noncentrosymmetric situation discussed in Refs. \cite{mockli2019,Mockli2020}.
When SOC, magnetic field, and ILH interplay, the $\eta_{00}$ even-parity singlets couple to the even-parity equal-spin field induced $\eta_{31(32)}$ triplets.

Since $\eta_{00}$ couples to $\eta_{31(32)}$, we also need the average
\begin{align}
\left\langle \hat{d}_{31(32)}f_{31(32)}\right\rangle_\mathbf{k} = 
\frac{|\omega_n|
\left[
 -\frac{iB\alpha}{\sqrt{2}}\eta_{00}+(\omega_n^2+B^2)\eta_{31(32)}
 \right]}{\left(\omega_n^2+t^2 \right )\left(\omega_n^2+B^2 \right )+\omega_n^2\alpha^2}.  
 \label{eq:f31}
\end{align}
This is obtained by writing the solution for $f_{31(32)}$ and then taking the average $\langle \hat{d}_{31(32)} f_{31(32)} \rangle_\mathbf{k}$; see Appendix \ref{app:low}.

We now feed Eqs. \eqref{eq:av_f00} and \eqref{eq:f31} to the self-consistency condition \eqref{eq:self_consistency}. Fixing $\eta_{00}$ to be real,
the resulting instability condition reveals that $\eta_{00}$ only couples to the imaginary part $\eta_{31(32)}=i\mathsf{Im}\,\eta_{31(32)}$, that is, the singlets and triplet order parameters have a relative phase difference of $\pi/2$. The self-consistency condition for the coupled $\{\eta_{00},\eta_{31(32)}\}$ state is
($T_{31}=T_{32}$)
\begin{align}
\begin{bmatrix}
\ln\frac{T}{T_{00}}+\mathsf{S}_{00} & \mathsf{S}_{00,31(32)}\\ 
\mathsf{S}_{00,31(32)} & 2\left(\ln\frac{T}{T_{31}}+\mathsf{S}_{31}\right)
\end{bmatrix}
\begin{bmatrix}
\eta_{00}\\ 
\mathsf{Im}\,\eta_{31(32)}
\end{bmatrix}
=0,
\label{eq:peta}
\end{align}
where
\begin{align}
&\mathsf{S}_{00} = \pi T\sum_{n\in\mathbb{Z}}\left[\frac{1}{|\omega_n|}-\frac{|\omega_n|\left(\omega_n^2+\alpha^2+t^2 \right )}{(\omega_n^2+t^2)(\omega_n^2+B^2)+\omega_n^2\alpha^2} \right ]; \notag \\
& \mathsf{S}_{31} = \pi T\sum_{n\in\mathbb{Z}}\left[\frac{1}{|\omega_n|}-\frac{|\omega_n|\left(\omega_n^2+B^ 2 \right )}{(\omega_n^2+t^2)(\omega_n^2+B^2)+\omega_n^2\alpha^2} \right ]; \notag \\
& \mathsf{S}_{00,31(32)} = \pi T\sum_{n\in\mathbb{Z}}\frac{\sqrt{2}B\alpha|\omega_n|}{(\omega_n^2+t^2)(\omega_n^2+B^2)+\omega_n^2\alpha^2} .
\label{eq:sums00}
\end{align}
Eq. \eqref{eq:sums00} clearly shows that the even-parity singlet-triplet coupling is a result of the joint action of magnetic field and SOC.
The three Matsubara sums can be evaluated as a sum of root functions. 
Rewriting the self-consistency Eq. \eqref{eq:peta} as $\mathsf{P}_\mathsf{even}\boldsymbol{\eta}=0$, the $(B,T)$ transition line is determined by $\det(\mathsf{P}_\mathsf{even})=0$.

In Fig. \ref{fig3:low} we show the transition lines for the combined even-parity $\{\eta_{00},\eta_{31(32)}\}$ state obtained from $\det(\mathsf{P}_\mathsf{even})=0$. In panels (a), (b) and (c) we vary $t$, $\alpha$ and $T_{31}$, respectively. Whereas $\alpha$ and the subleading triplet channel $T_{31}$ enhance the critical field, ILH $t$ makes the system more 3D, thus suppressing the critical field. Also, since $t$ appears in the denominator of Eq. \eqref{eq:sums00}, the larger $t$, the weaker the coupling between the order parameters. For $t=0$, the critical field is sensitive to the sub-leading triplet channel $T_{31}$ and diverges at low temperatures; see panel (a). 
The hoppings $t$ cut off the divergence.
The sensitivity to $T_{31}$ only remains as long as $t\lesssim T_{00}$.
In CeRh$_2$As$_2$, we expect $t>T_{00}$ such that $T_{31}$ looses relevance as a mechanism to enhance the upper critical field for the even-phase.

From this analysis, we conclude that for the low field even-phase, SOC provides the main mechanism for the enhancement of the critical field. Also, we can highlight that ILH and magnetic field work against the even-phase. The order parameter $\eta_{00}$ accommodates badly to the spin texture imposed by SOC together with ILH.

\subsection{The high-field odd-phase
\label{sec:high}
}

\begin{figure}
\includegraphics[width=0.65\linewidth, keepaspectratio]{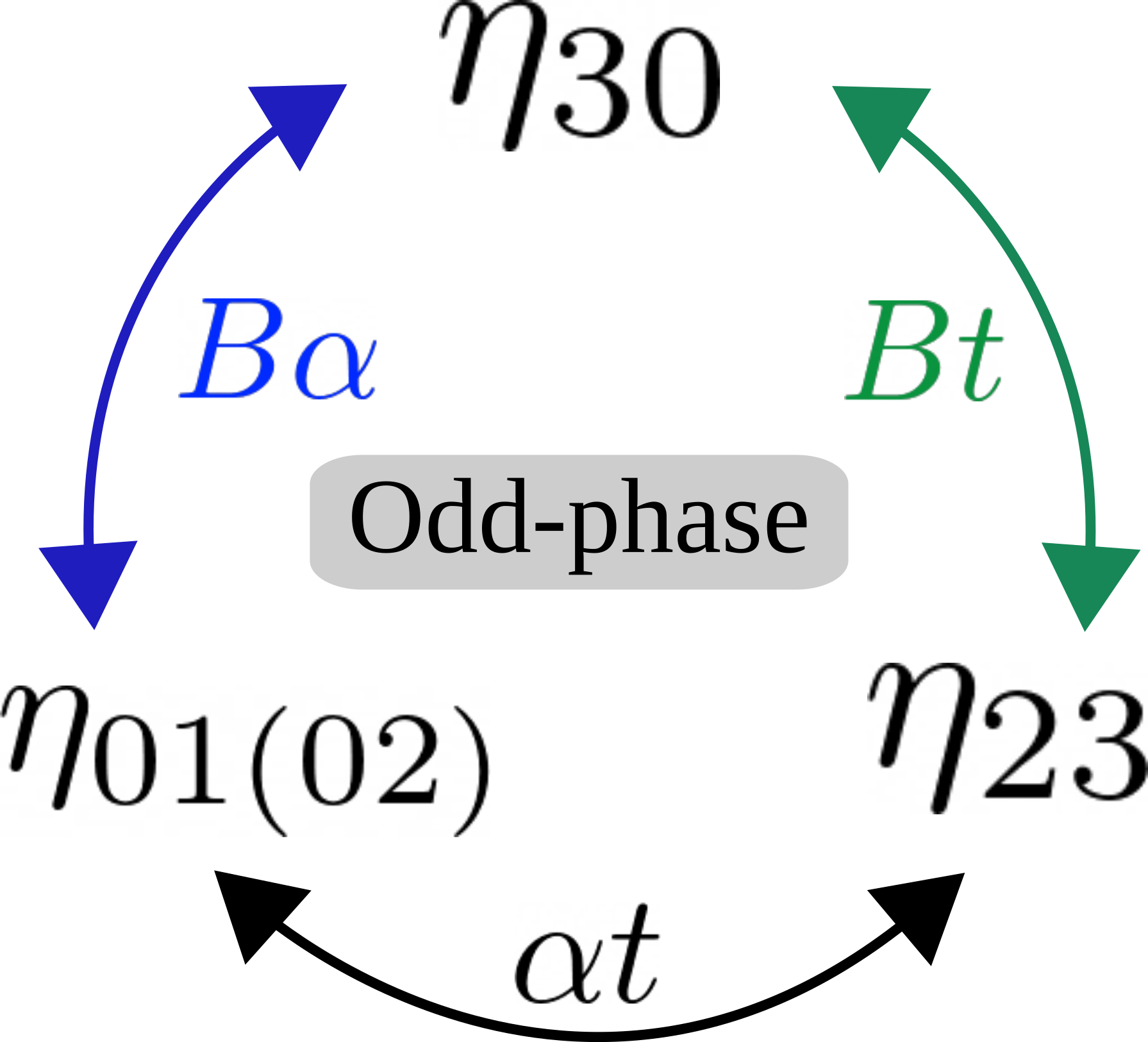}
\caption{
\label{fig:wheel}
Wheel showing that different pairwise combinations of the parameters $\{\alpha,t,B\}$ couple different odd-phase order parameters. At zero magnetic field, $\eta_{30}$ decouples from $\{\eta_{01(02)},\eta_{23}\}$.
SOC $\alpha$ couples $\{\eta_{01},\eta_{02}\}$. 
}
\end{figure}

\begin{figure*}
\includegraphics[width=\linewidth, keepaspectratio]{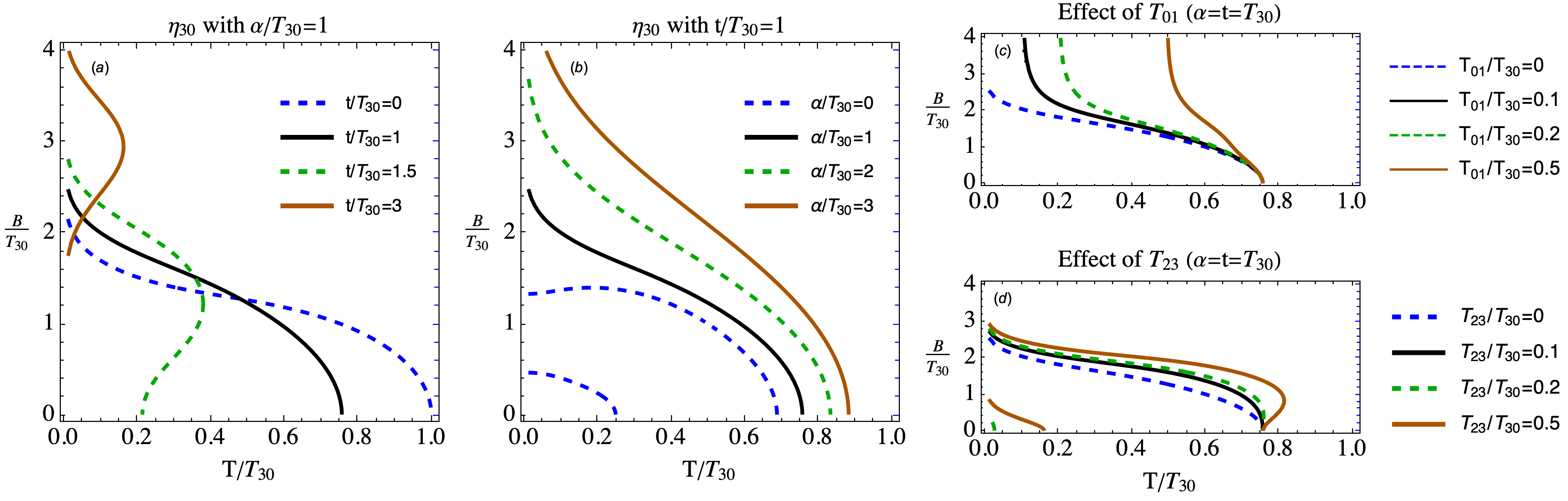}
\caption{
Transition lines of the high-field odd-phase.
(a) Reduction of the critical temperature at zero field and the enhancement of the critical field for larger ILH $t$. 
(b) Enhancement of both the critical temperature at zero field and the critical field for larger SOC $\alpha$. 
(c) Dominant $\eta_{30}$ phase taking into account the admixing of $\eta_{01(02)}$ triplets. In contrast to the even-phase, the odd-phase is sensitive to the subleading triplets.
(d) Dominant $\eta_{30}$ phase taking into account the admixing of $\eta_{23}$ triplets.
\label{fig:couplings}
}
\end{figure*}

In this section, we look at the odd solution of the linearized Eilenberger Eq. \eqref{eq:linEilenberger}, which has $\eta_{30}$ as the dominant order parameter.
Here, it is also possible to obtain analytic solutions if either $t_1$ or $t_2$ is included in the analysis.
For $t_1(t_2)$, 
we find a combined odd-parity $\{\eta_{30},\eta_{01},\eta_{02},\eta_{23}\}$ ($\{\eta_{30},\eta_{01},\eta_{02},\eta_{13}\}$) phase. 
Again, we use the properties listed in Fig. \ref{fig:tabela} together with the singlet-triplet coupling selected order parameter basis functions. The inter-sublattice order parameters $\eta_{23(13)}$ acquire the same basis function as the ILHs. 
Whether $t_1$ or $t_2$ is included is of secondary importance, and we choose to present the results for $t_1=t$.

We find the pair-breaking equation that describes the transition line of the coupled $\{\eta_{30},\mathsf{Im}\,\eta_{01},\mathsf{Im}\,\eta_{02},\mathsf{Im}\,\eta_{23}\}$ state. It reads
\small
\begin{align}
\det
\begin{bmatrix}
\ln\frac{T}{T_{30}}+\mathsf{S}_{30} & \mathsf{S}_{30,01} & \mathsf{S}_{30,02} & \mathsf{S}_{30,23} \\ 
\mathsf{S}_{30,01} & \ln\frac{T}{T_{01}}+\mathsf{S}_{01} & \mathsf{S}_{01,02} & \mathsf{S}_{01,23}  \\ 
\mathsf{S}_{30,02} & \mathsf{S}_{01,02} & \ln\frac{T}{T_{02}}+\mathsf{S}_{02} & \mathsf{S}_{02,23} \\ 
\mathsf{S}_{30,23} & \mathsf{S}_{01,23} & \mathsf{S}_{02,23} & \ln\frac{T}{T_{23}}+\mathsf{S}_{30} \\ 
\end{bmatrix}
=0.
\label{eq:pair_breakinh_high}
\end{align}
\normalsize
Let us also denote this condition by $\det(\mathsf{P}_\mathsf{odd})=0$.
We present the
detailed derivation of Eq. \eqref{eq:pair_breakinh_high} in Appendix \ref{app:high}.
The Matsubara sums are listed in Eqs. \eqref{eq:SA} -\eqref{eq:A1uA2u}, and analytically evaluated in Eqs. \eqref{eq:sa}-\eqref{eq:seg}. 
The order parameter $\mathsf{Im}\,\eta_{23}$ accompanies the same Matsubara sum as for $\eta_{30}$, which means that they would have the same phase diagram for a magnetic field along the $c$-axis if they were decoupled and had the same critical temperatures \cite{mockli2021scenarios}.
The term $\mathsf{S}_{30,23}$ causes their coupling, which, according to  Eq. \eqref{eq:A1uA2u}, happens only at finite magnetic field. 
The magnetic field induced $\eta_{01(02)}$ triplets couple to both $\eta_{30}$ and $\eta_{23}$, but in different ways. 
They couple to the $\eta_{23}$ triplets even without magnetic field, but only couple to $\eta_{30}$ at finite field. 
The wheel in Fig. \ref{fig:wheel} summarizes which pairwise combination of $\{\alpha,t,B\}$ couples which order parameters.

Starting the analysis considering that the critical temperature of the field induced triplets are negligible,  $T_{30}\gg T_{23},T_{01}$, we can discuss the effects of the normal state parameters; see Fig.\ref{fig:couplings}(a,b). We can conclude that a larger $\alpha$ enhances the Pauli limit, while a larger $t$ reduces the critical temperature at zero field. For strong enough ILH $t$, the transition line might develop a re-entrant  behavior. This is illustrated by the dashed-green and solid-brown curves in Fig. \ref{fig:couplings}a. If one looks at the experimentally obtained phase diagram of CeRh$_2$As$_2$, reproduced in Fig. \ref{fig:diagram}, the high-field transition line hints such a re-entrance. Figs. \ref{fig:couplings}(c,d), summarize the effects of the subleading triplet instabilities that couple to $\eta_{30}$ through the magnetic field. 
In panel (c) we show the effect of the $T_{01(02)}$ channel, and panel (c) shows the effect of adding the $T_{23}$ channel.

These results indicate that for the high-field odd-phase, both Rashba SOC and the presence of subleading instabilities provide mechanisms for the enhancement of the critical field. 
Whereas the even-phase is insensitive to its sub-leading channel $T_{31}$, the odd-phase is sensitive; compare Fig. \ref{fig3:low}c and \ref{fig:couplings}c. 
Therefore, if one finds evidence of an unconventional pairing mechanism that is able to stabilize the triplet channels in CeRh$_2$As$_2$, a naive fit to the experimental odd-phase transition line that does not take into account the sub-leading triplet channels might lead to inaccurate conclusions.  More precisely,  there is a surface in $\{\alpha,T_{01},T_{23}\}$ parameter space that yields the same $(T,B)$ point on the transition line. Additional knowledge about any of these parameters, either from experiments or first-principles calculations, could reduce the surface to a curve.

Mathematically, the insensitivity (sensitivity) of the even (odd) phase to $T_{31}$ ($T_{01}$) can be seen from the Matsubara sums $\mathsf{S}_{00,31(32)}$ in Eq. \eqref{eq:sums00} and $\mathsf{S}_{30,01(02)}$ in Eq. \eqref{eq:s3001}. In $\mathsf{S}_{00,31(32)}$, $t$ only occurs in the denominator, which shows that the larger $t$, the less relevant $T_{31}$. In $\mathsf{S}_{30,01(02)}$, due to the fact that $t$ occurs in both numerators and denominator has the consequence $T_{31}$ remains relevant. A similar analysis can be repeated for $T_{23}$.
Physically, in contrast to the odd-phase, ILH processes obstruct singlet to triplet conversion by magnetic field in the even-phase.

\subsection{Joining the even and odd phases}

\begin{figure}
\includegraphics[width=0.99\linewidth, keepaspectratio]{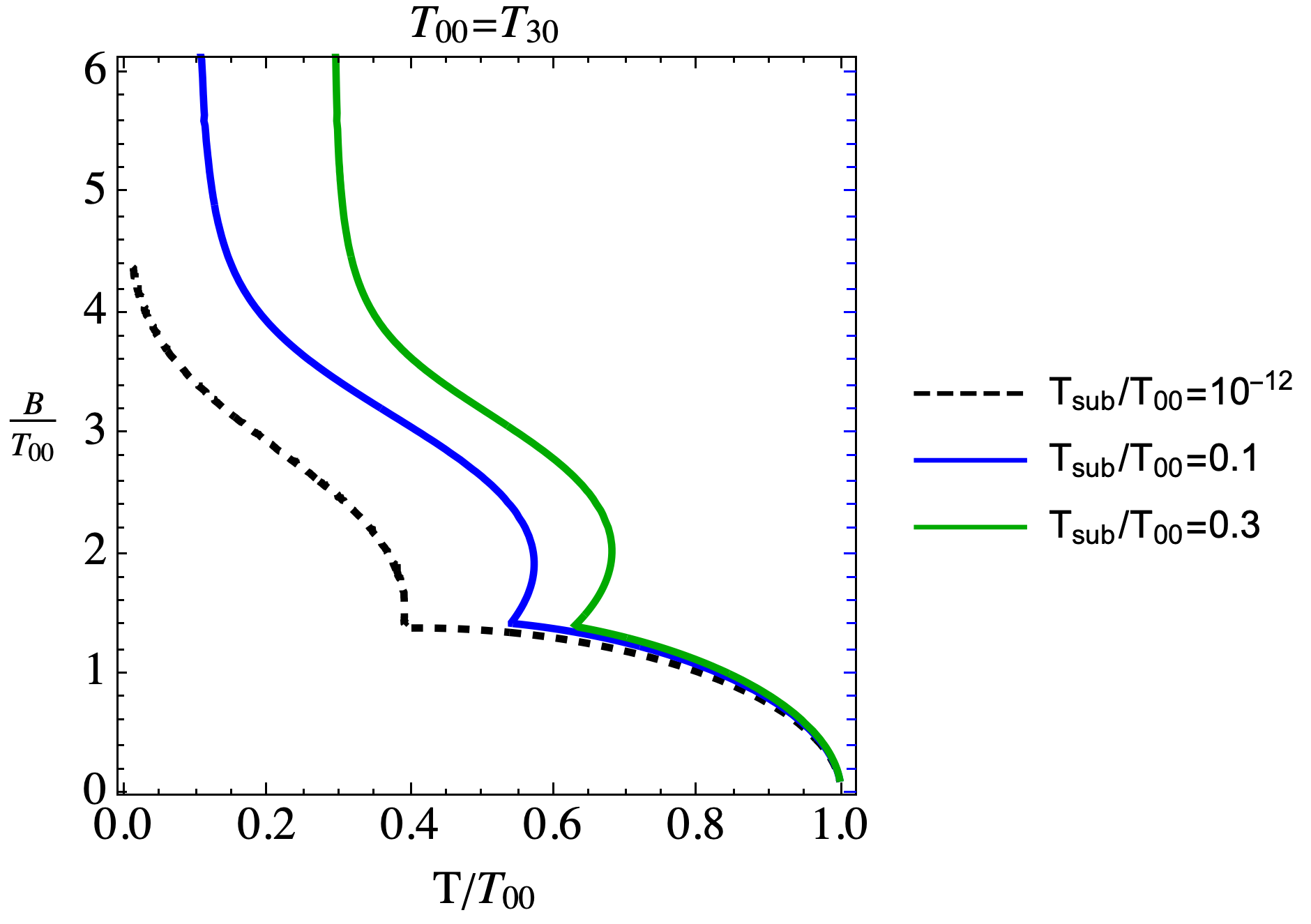}
\caption{
Full phase diagram with $t=\alpha=2T_{00}$ and $T_\mathsf{sub}=T_{31}=T_{01}=T_{23}$. The odd (even) phase is sensitive (insensitive) to the sub-leading triplet channels. A finite $T_\mathsf{sub}$ moves the multicritical point to the right and enhances the critical field. 
\label{fig:subdominant}
}
\end{figure}

We are now in a position to join the even and odd-phase solutions into a single phase diagram. Because the even and odd order parameters do not couple at the linearized level, 
we know that they are mutually excluding phases. 
Another way to see this is that one can not continuously deform $\eta_{00}$ into $\eta_{30}$, which guarantees a fist-order phase transition between them \footnote{Assuming that there is no additional spontaneous symmetry broken phase.}. 
The second-order transition line including both phases can be obtained by a simple comparison of the even and odd-phase instability conditions. The realized transition line is determined by $\det(\mathsf{P})=0$, where
\begin{align}
\begin{cases}
\mathsf{P}=\mathsf{P}_\mathsf{even}\,\,\mbox{if}\,\, \det(\mathsf{P}_\mathsf{even})<\det(\mathsf{P}_\mathsf{odd}) \\
\mathsf{P}=\mathsf{P}_\mathsf{odd}\,\,\mbox{otherwise.}
\end{cases}
\label{eq:minumum}
\end{align}
An alternative way to see this is noticing that the superconducting free energy depends on the determinants \cite{Coleman2015}. Eq. \eqref{eq:minumum} determines the lower free energies between the even and odd phases. 
Here, we do not obtain the first order transition line, as this would require a treatment beyond linearization. 

In Fig. \ref{fig:subdominant} we illustrate the superconducting transition as determined by Eq. \eqref{eq:minumum}. The figure also shows the insensitivity (sensitivity) of the even (odd) phase to its subleading instabilities. For illustrative purposes, we set all sub-leading critical temperatures to a single value $T_\mathsf{sub}$. The presence of $T_\mathsf{sub}$ moves the multicritical point to the right.

\section{Effect of scalar impurities \label{sec:impurities}}

Here, we address the effect of isotropic scalar impurities introduced in Eq. \eqref{eq:impurity_selfenergy} on the phase diagram.
The question that motivates this section is: does the system respond to disorder like a centrosymmetric or a noncentrosymmetric superconductor? 
The short answer is neither. 
We now show that for the even-phase, the response depends on the energy scale ratio $\alpha/t$. 
The odd-phase has its own peculiar behavior that
depends on both $\alpha/t$ and $t/T_c$.   As a  minimal model calculation, below we only consider the dominant  $\eta_{00}$ and $\eta_{30}$ order parameters.

\begin{figure*}
\includegraphics[width=\linewidth, keepaspectratio]{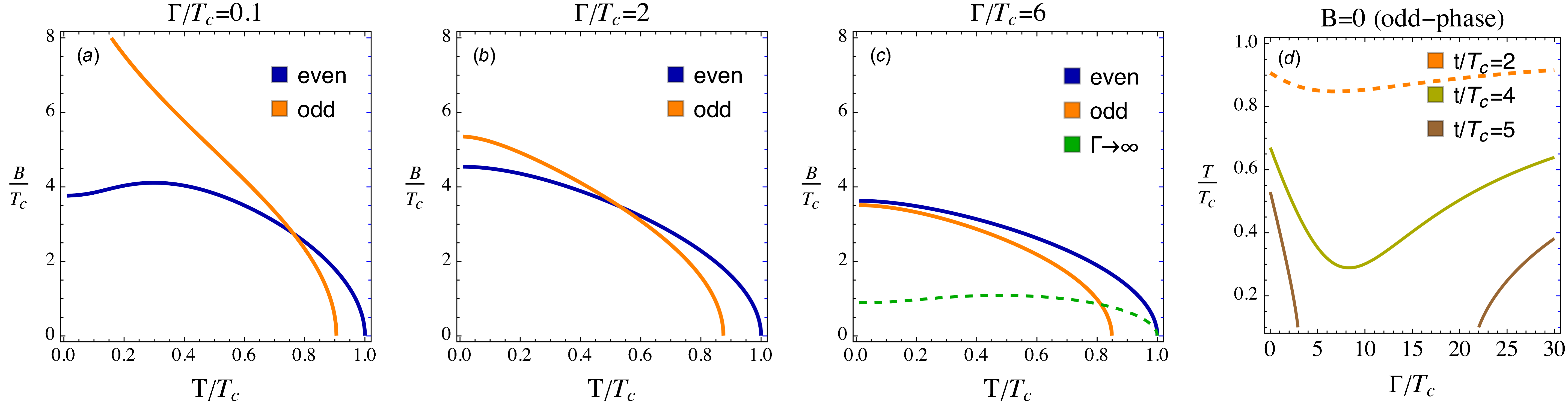}
\caption{
\label{fig:disorder}
The effect of impurities on the even and odd phases. We set $T_{00}=T_{30}=T_c$, $t/T_c=2$ and $\alpha/t=5$.
(a) Almost clean situation with $\Gamma/T_c=0.1$.
(b) $\Gamma/T_c=2$. The odd-phase state suffers more than the even-phase. Nonetheless, the odd-phase is more robust than a naive unconventional state which is obliterated for $\Gamma/T_c\sim 1$. 
(c)  $\Gamma/T_c=6$. The odd-phase is now subdominant with respect to the even-phase. 
(d) The effect of the scattering rate on the effective zero field odd-phase transition temperature $T_{30}^*$. If $t$ is sufficiently small (up to $t/T_c=4$), the pair-breaking effect by impurities is continuous and non-monotonic. For higher values of $t$ ($t/T_c=5$), pair-breaking becomes qualitatively Abrikosov-Gor'kov like. 
$T_{30}^*$ of the odd-phase might revive in dirty systems, but remains subdominant with respect to the even-phase.
}
\end{figure*}

\subsection{Disordered even-phase}

For the even-phase, we solve for the corresponding correlations $f_{00}$; see Eq. \eqref{eq:f00}. Note that even if $d_{00}$ is enforced to be the only allowed order parameter in Eq. \eqref{eq:f00}, other $\bar{d}_{ab}$'s are populated by the impurities. Taking the average of Eq. \eqref{eq:f00} one sees that the only other average that contributes is $\bar{d}_{03}=\Gamma\langle f_{03}(\mathbf{k};\omega_n)\rangle_\mathbf{k}$. 
Solving the Eilenberger equation for $\langle f_{03}\rangle_\mathbf{k}$, we obtain the odd-frequency pairing correlation average 
\begin{align}
\langle f_{03}\rangle_\mathbf{k} = \frac{iB\mathsf{sgn}(\omega_n)\left[\bar{\omega}_n^2+t^2\right]\eta_{00}}{|\omega_n||\bar{\omega}_n|\left(|\omega_n||\bar{\omega}_n|+\alpha^2\right)+\omega_n^2t^2+B^2\left(\bar{\omega}_n^2+t^2\right)}. \label{eq:odd_frequency}
\end{align}
Substituting Eq. \eqref{eq:odd_frequency} into Eq. \eqref{eq:f00} and taking the average, we then obtain 
\begin{align}
\langle f_{00}\rangle_\mathbf{k} = \frac{\left[|\bar{\omega}_n|\left(|\omega_n||\bar{\omega}_n|+\alpha^2\right)+|\omega_n| t^2\right]\eta_{00}}{|\omega_n||\bar{\omega}_n|\left(|\omega_n||\bar{\omega}_n|+\alpha^2\right)+\omega_n^2t^2+B^2\left(\bar{\omega}_n^2+t^2\right)}.
\label{eq:avf00}
\end{align}
Note that for $B=0$, this simply yields $\langle f_{00}\rangle_\mathbf{k}=\eta_{00}/|\omega_n|$, which shows that the critical temperature  remains unaffected in the presence of impurities, as guaranteed by Anderson's theorem \cite{parks1969vol2}.
Also, for $\alpha=0$ we have $\langle f_{00}\rangle_\mathbf{k}=|\omega_n|\eta_{00}/(\omega_n^2+B^2)$, from which one obtains the Pauli limiting effect.  Eq. \eqref{eq:avf00} is fed to the self-consistency condition Eq. \eqref{eq:self_consistency} that after performing the Matsubara sum yields now a $\Gamma$ dependent transition line. 

An increasing scattering rate $\Gamma$ undoes not only the critical field enhancement by SOC, but also the effects related to ILH. In Fig. \ref{fig:disorder}, the effect of $\Gamma$ on the even transition line is illustrated by the blue curves. 
In the plots, $t/T_c=2$ and $\alpha/t=5$. 
Panel (a) shows an almost clean situation with $\Gamma/T_c=0.1$. At low temperatures, one can identify a $t$ caused depression, which cuts off the critical field even in the clean case. As discussed in Sec. \ref{sec:normal_state}, we expect $\alpha>t$ in CeRh$_2$As$_2$, so  impurities  first undo the effects due to $t$. Comparing panels (a) and (b), the $t$ caused depression is undone by $\Gamma$. 
Next, a stronger scattering rate of $\Gamma/T_c=6$ starts undoing the critical field enhancement caused by $\alpha$. Comparing panels (b) and (c), the  critical field is reduced. In the $\Gamma\rightarrow\infty$ limit, the transition line is set by the Pauli limit; see dashed-green curve in panel (c). 
Remarkably, as impurities first undo the critical field suppression caused by $t$, they effectively generate a critical field enhancement. Therefore,  one can expect larger critical fields in the even-phase for disordered samples. 
We also illustrate this in unusual behavior in Fig. \ref{fig:full}, and by an animated version of Fig. \ref{fig:disorder} found in the online supplemental material \footnote{See Supplemental Material at [URL will be inserted by publisher] for a detailed derivation \label{foot}}.

\subsection{Disordered odd-phase}

We repeat an analogous procedure for $f_{30}$. Now, the impurities populate $\bar{d}_{33}=\Gamma\langle f_{33}(\mathbf{k};\omega_n)\rangle_\mathbf{k}$. Solving for $\langle f_{33}\rangle_\mathbf{k}$, then substituting into the solution for $f_{30}$ and taking the average, we obtain 
\begin{widetext}
\small
\begin{align}
\langle f_{30}\rangle_\mathbf{k} =  \frac{\left[\left(\bar{\omega}_n^2+\alpha^2 \right )\left(|\omega_n||\bar{\omega}_n|+\alpha^2+t^2 \right )+B^2\left(|\omega_n||\bar{\omega}_n|+\alpha^2 \right ) \right ]\eta_{30}}{\left(|\omega_n||\bar{\omega}_n|+\alpha^2+t^2 \right )\left[|\omega_n| \left(\bar{\omega}_n^2+\alpha^2\right)+|\bar{\omega}_n|t^2\right ]+B^2\left[2|\omega_n|\left(\omega_n^2+\alpha^2 \right )+\Gamma\left(\alpha^2+4\omega_n^2 \right )+\Gamma^2\left(3|\omega_n|+\Gamma \right )-2|\bar{\omega}_n|t^2 \right ]+B^4|\bar{\omega}_n|}. 
\label{eq:avf30}
\end{align}
\end{widetext}
Substitution into the self-consistency condition Eq. \eqref{eq:self_consistency} yields the pair-breaking equation in terms of a Matsubara sum that can be expressed in terms of root functions for efficient plotting purposes.

Let us examine the clean and dirty limits.
In the clean limit ($\Gamma=0$), we recover (see Appendix \ref{app:high})
\begin{align}
\langle f_{30}\rangle
_\mathbf{k}=
\frac{\left(\omega_n^2+\alpha^2\right)\left(\omega_n^2+\alpha^2+B^2+t^2\right)\eta_{30}}{|\omega_n|\left[\omega_n^2+\alpha^2+\left(B+t\right)^2\right]\left[\omega_n^2+\alpha^2+\left(B-t\right)^2\right]}. \notag 
\end{align}
The opposite $\Gamma\rightarrow\infty$ limit gives
$\langle f_{30}\rangle_\mathbf{k}=|\omega_n|\eta_{30}/(\omega_n^2+B^2)$, and the transition line is determined simply by the Pauli limit.

\begin{figure}
\includegraphics[width=0.97\linewidth, keepaspectratio]{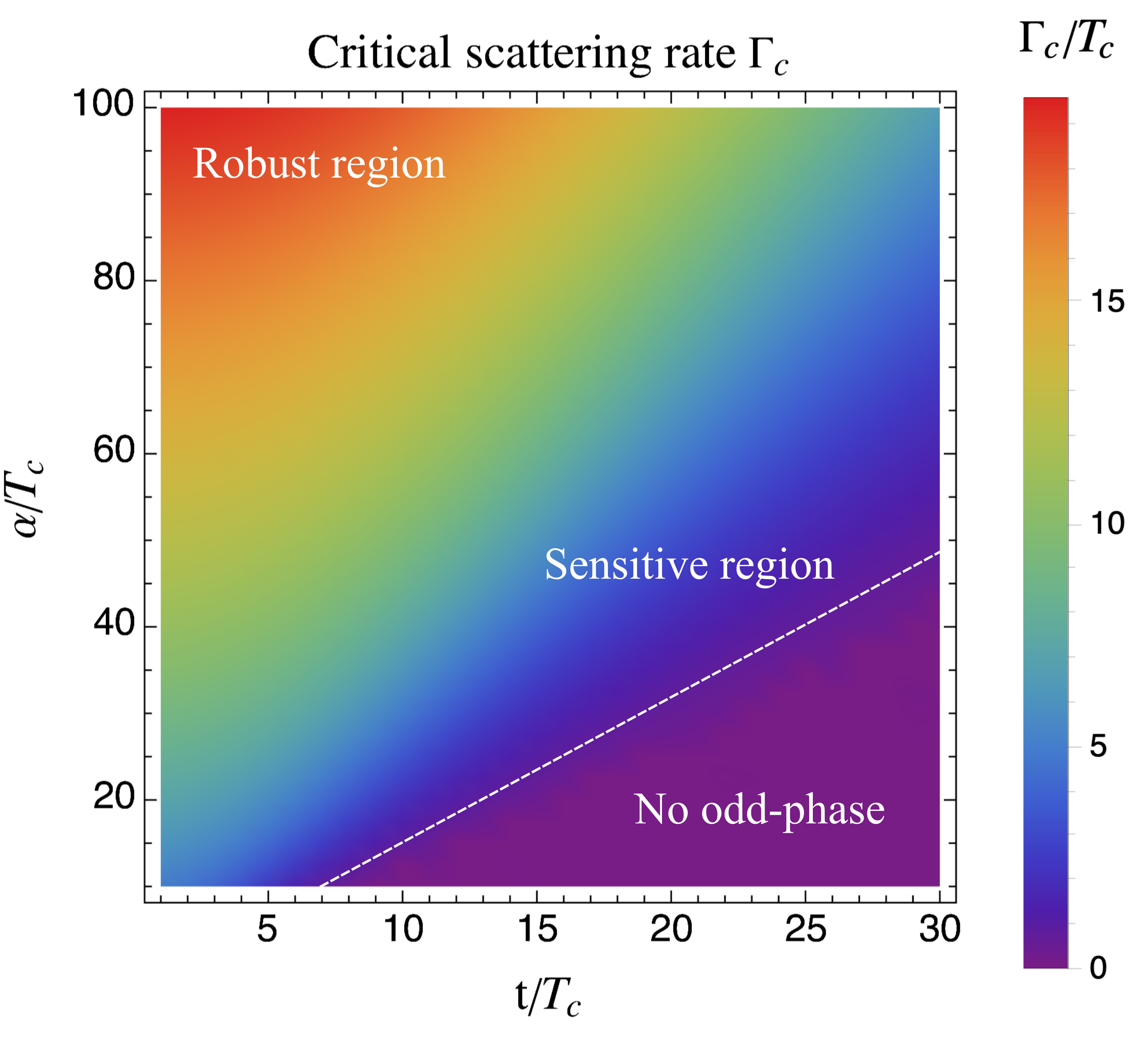}
\caption{
\label{fig:critical_scattering}
Color map of the critical scattering rate $\Gamma_c$ necessary to obliterate the odd-phase as a function of $\alpha$ and $t$. 
There is no odd-phase in the purple region. The purple-blue border delimits the region where the odd-phase is sensitive $\Gamma_c/T_c\sim 1$. 
The odd-phase is robust in the region $\alpha/t>1$ with $\Gamma_c/T_c\gtrsim 10$.
}
\end{figure}

Since for the large $\Gamma$ limit the even and odd solutions merge, one can define a critical impurity scattering rate $\Gamma_c$ for which the odd-phase becomes subdominant with respect to the even-phase; see Fig. \ref{fig:disorder}(c).
In Fig. \ref{fig:critical_scattering} we show a color map of the critical scattering rate $\Gamma_c$ as a function of SOC and ILH. We identify three regions of the parameter space: (i) no odd-phase, which corresponds to the purple region; (ii) the purple-blue border at which $\Gamma_c/T_c\sim 1$ corresponding to a sensitive odd-phase; (iii) the green to red region with $\alpha/T_c>1$ for which the odd-phase is robust $\Gamma_c/T_c\gtrsim 10$.

\section{Discussion \label{sec:discussion}}

Based on our results, we now discuss the phase diagram of CeRh$_2$As$_2$. We start revising the upper critical fields as measured by experiment \cite{khim2021fieldinduced}, from which we can extract a set of parameters that can quantitatively reproduce the phase diagram. Then, we contrast these parameters with what is known from first principles calculations and compare CeRh$_2$As$_2$ to other related materials. Furthermore, we speculate on future developments in both theory and experiment.

\subsection{Experimentally observed upper critical field}

CeRh$_2$As$_2$ displays two superconducting phases as a function of $c$-axis magnetic field (see Fig. \ref{fig:diagram}). The low field even-phase is associated with the (extrapolated) upper critical field of about 5T, while the high-field odd-phase is robust up to about 15T. Under in-plane magnetic fields, only the even-phase is observed, and it survives up to magnetic fields of the order of 2T. All the critical fields mentioned above are remarkably high for a material with a critical temperature of about 0.26K.

To get a more quantitative measure of these unusually high upper critical fields, we can think in terms of the Pauli limit $B_\mathsf{P}$. For  weakly coupled superconductors without SOC, the Pauli limit is given by \cite{saint1969type}
\begin{align}
    \mu_\mathsf{B}B_\mathsf{P} = \frac{\sqrt{2}\Delta_0}{g}= \frac{\pi\sqrt{2}}{e^\gamma}\frac{k_\mathsf{B}T_c}{g}.
\end{align}
For $g=2$, the ratio $ \mu_\mathsf{B}B_\mathsf{P} /(k_\mathsf{B}T_c)\approx 1.25$. 
In CeRh$_2$As$_2$, the in-plane ($\parallel$) and perpendicular ($\perp$) $g$-factor is estimated to be $g_\parallel\approx1.43$ and $g_\perp \approx1.11$ \cite{khim2021fieldinduced}. This leads to the naive ratio estimates of  $ \mu_\mathsf{B}B_\mathsf{P}^\parallel /(k_\mathsf{B}T_c)\approx 1.74$ and $ \mu_\mathsf{B}B_\mathsf{P}^\perp /(k_\mathsf{B}T_c)\approx 2.25$. 
The experimentally obtained upper critical fields in CeRh$_2$As$_2$ display the ratios $ \mu_\mathsf{B}B_\mathsf{c2 Low}^\parallel /(k_\mathsf{B}T_c)\approx 5$, $ \mu_\mathsf{B}B_\mathsf{c2 Low}^\perp /(k_\mathsf{B}T_c)\approx 12$, and $ \mu_\mathsf{B}B_\mathsf{c2 High}^\perp /(k_\mathsf{B}T_c)\approx 36$ \cite{khim2021fieldinduced}.
All upper critical fields are beyond the Pauli limit.
Therefore, it is interesting to introduce the measure of a Pauli limit violation ratio  defined as $\mathsf{PVR} = B_\mathsf{c2}/B_\mathsf{P}$ \cite{Cao:2021}, which take values $\mathsf{PVR}_\mathsf{Low}^\parallel \approx 3$, $\mathsf{PVR}_\mathsf{Low}^\perp \approx 5$, and $\mathsf{PVR}_\mathsf{High}^\perp \approx 16$ \footnote{The $\mathsf{PVR}$ value here should not be confused with the Maki value $\alpha_\mathsf{M}=\sqrt{2}B_\mathsf{c2}^\mathsf{theory}/B_\mathsf{P}$ \cite{Matsuda2007}, where $B_\mathsf{c2}^\mathsf{theory}=\Phi_0/(2\pi\xi_0^2)$ is the theoretical orbital limit. 
The numerator for the $\mathsf{PVR}$ value is taken directly from the experiment in Ref. \cite{khim2021fieldinduced}.}.

The crystal structure of CeRh$_2$As$_2$ (see Sec. \ref{sec:model}) reveals that the small in-plane Pauli limit enhancement might be attributed to an Ising SOC component $\lambda$, whereas the larger enhancement of the perpendicular Pauli limit stems from the Rashba SOC component \cite{khim2021fieldinduced,mockli2021scenarios}. We comment on the constraints on parameter regimes for both magnetic field directions separately below.

\subsection{Enhanced c-axis Pauli limit}

\subsubsection{Low-field phase}

From Section \ref{sec:low}, we conclude that for the low-field phase, the upper critical field can only be enhanced by a finite Rashba SOC. 
For a $\mathsf{PVR}_\mathsf{Low}^\perp \approx 5$, our analysis suggests a ratio $\alpha/t \approx 5$ for a relatively disordered superconductor with $\Gamma/(k_\mathsf{B}T_c)\approx 4$ (See full green line if Fig. \ref{fig:full}). 
Interestingly, 
a theoretical fit to the experimental phase diagram (ignoring orbital limiting) is only obtained in the presence of impurities; compare solid and dashed curves in Fig. \ref{fig:full}.

\subsubsection{High-field phase}

From Section \ref{sec:high}, we concluded that both Rashba SOC and the presence of subleading instabilities are good mechanisms for the enhancement of the upper critical field in the high-field phase. 
For this phase, ILH $t$ suppresses the critical temperature of the high-field phase. 
The (extrapolated) value of the critical temperature for the high-field phase is $\approx 0.16$K, with a reentrance of about $0.02$K.
Assuming the intra-layer pairing mechanism to be the same for both low and high-field phases, our analysis suggest $t/T_c\approx 25$.

\begin{figure}
\includegraphics[width=0.99\linewidth, keepaspectratio]{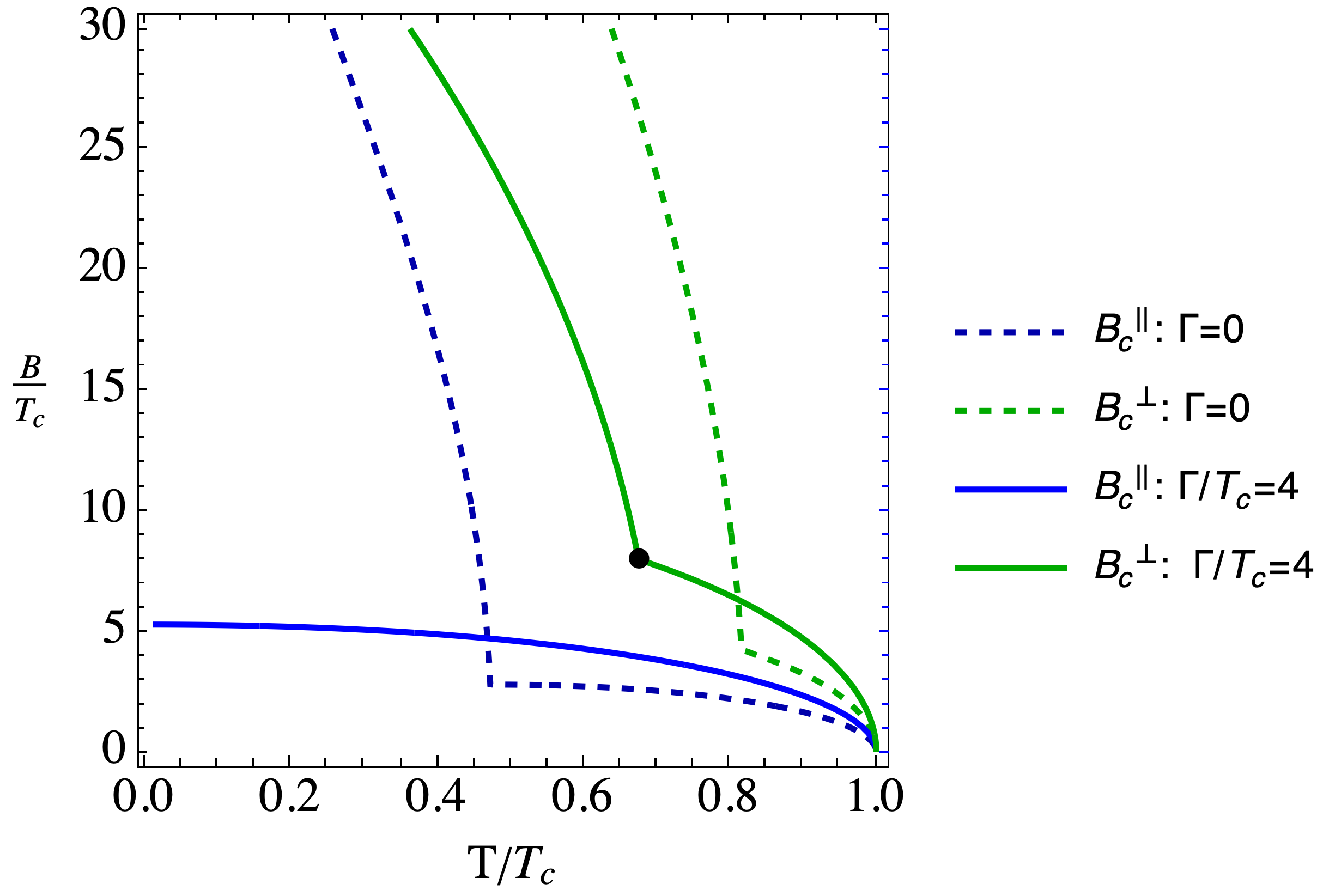}
\caption{
\label{fig:full}
Transition lines comparing the clean (dashed) and disordered cases (solid).
The in-plane ($\parallel$) lines were obtained by considering only the Ising SOC component $\lambda$, and the perpendicular ($\perp$) by considering only Rashba SOC $\alpha$. We used $t/T_c=25$, $\alpha /t=5$ and $\lambda /t=2.4$. For these parameters, the perpendicular odd-phase becomes subdominant to the even-phase for $\Gamma/T_c\gtrsim 12.8 $. 
The disorder enhances the even-phase by undoing the suppression caused by $t$.
The black point indicates the thermodynamic multicritical point. 
}
\end{figure}

Concerning the magnitude of SOC, if we neglect the effects of subleading instabilities and use the value of $\alpha/t\approx 5$ that is in agreement with the low-field phase, we find a $\mathsf{PVR}\approx 16$, in good agreement with what is observed for the high-field phase (see solid green line in Fig. \ref{fig:full}). This shows that the phase diagram can be quantitatively reproduced, considering only the effects of Rashba SOC (and impurities). 
It is important to emphasize here, though, that a similar phase diagram could be obtained for a finite $T_\mathsf{sub}$ and different ratios of $t/T_c$, $\alpha/t$, and $\Gamma/T_c$. Note, also, that the consideration of orbital depairing would give even more freedom for the fitting parametrization parameters.

\subsection{Enhanced in-plane Pauli limit}

Up to now, we discussed the pair-breaking equations for a 
These equations remain invariant as long as the applied magnetic field is perpendicular to the SOC texture. This allows us to qualitatively assess the in-plane magnetic field enhancement caused by perpendicular Ising SOC. The same solutions apply, but now $B$ is in-plane and $\alpha\rightarrow\lambda$. From the discussion in Section \ref{sec:low}, we conclude that for the low-field phase in presence of in-plane magnetic fields, the critical field can only be enhanced by a finite Ising SOC. 
For a $\mathsf{PVR}\approx3$, our analysis suggest a ratio $\lambda/t \approx 2.4$, fixing $\lambda \approx 1.2$meV.

\subsection{Effects of impurities}

\subsubsection{Low-field phase}

The critical temperature $T_{00}$ is unaffected by non-magnetic impurities if the pairing is of $s$-wave nature, as expected by Anderson's theorem. 
Note, though, that impurities lead to non-trivial effects on the critical field in both field directions. 
In particular, for scattering rates $\Gamma$ satisfying $T_c<\Gamma\lesssim t$, the detrimental effect of ILH on the critical field is reduced, which effectively leads to an enhancement of the critical field in disordered samples.
For strong scattering rates $t<\Gamma\lesssim \alpha,\lambda$, impurities would start undoing the enhancement promoted by Rashba or Ising SOC, leading to a reduction of the critical field in more disordered samples. This effect is clearly seen in Fig. \ref{fig:disorder}.  This non-monotonic dependence of the critical field as a function of the impurity scattering rate seems to be a unique feature of locally noncentrosymmetric superconductors.

\subsubsection{High-field phase}

The critical temperature $T_{30}$ is suppressed by non-magnetic impurities (take $B=0$ in Eq. \eqref{eq:avf30}).  This can be understood by the unconventional nature of this order parameter, which has a staggered phase across the layers. Even if pairing is of $s$-wave nature within the layers, a finite ILH introduces horizontal line nodes, making this superconducting state susceptible to impurities. This phenomenon can be understood in terms of a generalized Anderson's theorem \cite{Andersen:2020}.

\subsubsection{Critical scattering rate and multicritical point}

Putting the low and high-field phases together, we define a critical scattering rate $\Gamma_c$ above which the high-field phase becomes subdominant with respect to the low-field phase at all temperatures and fields. This maximum scattering rate depends on both $\alpha/T_c$ and on $t/T_c$. 
Given the parameters discussed above, the current samples of CeRh$_2$As$_2$ are located in the robust region depicted in Fig. \ref{fig:critical_scattering}, with the high-field phase been at least ten times more robust against impurities than one would expect for an unconventional superconducting state with nodes. 
For the parameters used in Fig. \ref{fig:full}, a scattering rate of $\Gamma/T_c\approx 12.8$ would be necessary to obliterate the odd-phase.

Impurities also affect the position of the multicritical point in the $(B,T)$ phase diagram.
In the clean case, larger values of $\alpha/t$, $T_\mathsf{sub}$ and $T_{30}$ move the multicritical point to the right. 
Assuming that $\alpha>t$, a larger scattering rate $\Gamma$ affects the multicritical point in two ways: (i) $\Gamma$ enhances the critical field of the even-phase moving the multicritical point up; (ii) $\Gamma$ suppresses the odd-phase's critical temperature $T_{30}$, which moves the multicritical point to the left.

\subsubsection{Absence of the odd-phase for in-plane fields}

Interestingly, in the clean limit, an in-plane magnetic field together with the Ising SOC also favors the odd-phase; see dashed-blue curve in Fig. \ref{fig:full}. However, because $\lambda<\alpha$, disorder and ILH are more effective in suppressing the odd-phase for in-plane fields. For the parameters used in Fig. \ref{fig:full}, the in-plane odd-phase disappears for $\Gamma_c/T_c\gtrsim 2.9$.
Note that Fig. \ref{fig:critical_scattering} can be reinterpreted for $\lambda$ in an in-plane magnetic field. 
This means that the in-plane high-field phase is at the border of the sensitive region. This explains why the high-field phase is not observed in current experiments for in-plane fields, and suggests that its observation might be possible in cleaner samples. A systematic experimental study of samples with different residual sensitivities and the corresponding variations in the upper critical field of the odd-phase could assess this behavior. 
Since for CeRh$_2$As$_2$ $k_\mathsf{B}T_c\approx 0.02$ meV, a scattering rate of the same order is expected to correspond to a rather clean system. 
Controlling the amount of disorder by external means, such as by electron irradiation \cite{Cho2018,Timmons:2020}, could provide a interesting line for further experimental investigations.

\subsection{Hierarchy of energy scales}

From the discussion above, we conclude that only if $\alpha>\lambda>t$ is assumed, one can generate a phase diagram in qualitative agreement with  experiments. For a quantitative agreement, we use $t/T_c=25$, $\alpha /t=5$, $\lambda /t=2.4$, and $\Gamma/T_c = 4$ to generate the phase diagram displayed in Fig. \ref{fig:full}. 

The energy scale hierarchy $\alpha>\lambda>t$ might seem non-intuitive at first. Note that the ILH amplitude $t$ is dominated by hopping processes between nearest layers, while $\lambda$ is associated with Ising SOC originated from next-nearest layer processes. In this context, we need to remind the reader that these parameters enter the quasiclassical formalism as effective contributions of the corresponding terms at the Fermi surface. These parameters are all accompanied by non-trivial momentum dependent form factors, as discussed in section \ref{sec:normal_state}. Interestingly, depending of the position of the Fermi surfaces on the Brillouin zone, this non-intuitive hierarchy can be satisfied, as recently discussed in \cite{Cavanagh:2021}.

We briefly discuss how this energy scales hierarchy is in agreement with recent first-principles calculations \cite{ptok2021electronic,nogaki2021topological}. Looking at the dispersion of the bands crossing the Fermi surface along the $\Gamma-Z$ direction, we can estimate an ILH amplitude of the order of 100 meV. With $k_\mathsf{B} T_c \approx 0.02$ meV, the condition $t/T_c \gg 1$ is satisfied and we can safely neglect the presence of subleading instabilities  for the enhancement of the upper critical field of the low-field phase. The estimate of Rashba and Ising SOCs from these calculations is not so straightforward. Momentum-dependent SOC can be estimated from a comparison of the bands calculated in presence and absence of relativistic effects, as presented in Fig. S4 of Ref. \cite{ptok2021electronic}. For bands far away from the Fermi surface there are clear momentum dependent band splittings of the order of 20 0meV (in particular around -2 eV along the $X-M$ direction). Note, though, that the electrons associated with these bands are primarily d-electrons from Rh atoms, so we expect the magnitude of these momentum-dependent SOC terms in the bands closer to the Fermi surface to be larger, since these bands are composed primarily of f-electrons from the Ce atoms, with larger atomic number. This simple analysis suggests that SOC terms can be in fact larger than ILH in this material. 

\subsection{Comparison to other materials}

Phase diagrams with multiple superconducting phases are rare, and have only been reported in other two heavy fermion materials: UPt$_3$ \cite{Fischer:1989,Adenwalla:1990,Joynt:2002} and  UTe$_2$ \cite{Ran:2019}. The presence of strong correlations or quantum fluctuations supporting pairing in unconventional superconducting channels seems to be a key ingredient among these materials \cite{Dyke:2014,White:2015,Landaeta:2018,Smidman:2018}.

It is worth noting that there are several superconducting materials with the same CaBe$_2$Ge$_2$-type structure: LaIr$_2$Si$_2$ \cite{Braun:1983}, SrPd$_2$Bi$_2$ \cite{Xie:2016}, SrPd$_2$Sb$_2$ \cite{Kase:2016b}, SrPt$_2$As$_2$  \cite{Kase:2016a}, LaPd$_2$Bi$_2$ \cite{Mu:2018}, and LaPd$_2$Sb$_2$ \cite{Ganesanpotti:2014}. All these materials have a superconducting critical temperature of around 1K, but none displays the remarkable transition to a high-field phase within the superconducting state. Interestingly, these materials usually display two polymorphs, the tetragonal form with CaBe$_2$Ge$_2$-type structure, and the monoclinic form with ThCr$_2$Si$_2$-type structure, the first consistently favoring the occurrence of superconductivity  \cite{Xie:2016,Zheng:1986,Shelton:1984}. One possible explanation for this distinction is the proximity to a van Hove singularity for the CaBe$_2$Ge$_2$-type materials, which is often associated with a nearby structural, electronic or magnetic instability \cite{Xie:2016, ptok2021electronic}. A systematic study of materials in the pnictide family suggests that phonon-mediated pairing provides a consistent picture for the origin of superconductivity in these systems \cite{Kase:2016b}. The development of conventional superconductivity in these isostructural materials suggests that the lack of inversion symmetry alone does not guarantee the development of an unconventional superconducting phase at high magnetic fields. 

One fundamental aspect that distinguishes CeRh$_2$As$_2$ from the isostructural materials mentioned in the previous paragraph is the presence of localized f-electrons in the Ce atoms. CeRh$_2$As$_2$ can be thought of as a Kondo lattice material, with one localized f-electron associated with each Ce$^{3+}$ ion coexisting with a sea of light conduction electrons stemming from the Rh and As atoms. At high temperatures, the localized f-electrons act as incoherent scattering centers, but below the coherence temperature, $T_\mathsf{coh}$,  the localized and itinerant electrons hybridize, giving rise to a coherent heavy Fermi liquid with an enhanced effective mass \cite{Coleman:2007,Coleman:2015}. 
For CeRh$_2$As$_2$, the characteristic maximum of the resistivity as a function of temperature suggests $T_\mathsf{coh} \approx 20-40$K. At 0.5K, the specific heat coefficient reaches values of the order of 1 J/mol K$^2$. These two observations indicate the presence of well-defined heavy quasiparticles at the Fermi surface just above the superconducting state \cite{khim2021fieldinduced}. The heavy fermion nature of the normal state can be important in two different ways. First, the magnitude of the Rashba SOC depends on the atomic number $Z$, therefore CeRh$_2$As$_2$ should display the largest Rashba SOC among the mentioned materials. Second, the heavy fermion character might be important to guarantee a mechanism for superconductivity in unconventional channels associated with the subleading instabilities discussed here.

\subsection{Perspectives}

Our formalism introduces the effects of magnetic field by directly coupling to the spin degree of freedom. 
In order to consider also the orbital effect, one might develop a microscopic Ginzburg-Landau expansion, for which our work provides a suitable starting point.
One could for instance derive the Ginzburg-Landau coefficients as used in Ref. \cite{schertenleib2021unusual} from the microscopic theory provided here. 
For a more realistic treatment, we suggest starting from the Eilenberger matrix equation \eqref{eq:eilenberger_matrix},  
consider $\alpha$, $\lambda$, $t_1$ and $t_2$ simultaneously,
and develop the corresponding Riccati equations for numerical simulations of the Abrikosov lattice. 
This would allow for a realistic study of the vortex state throughout the first-order phase transition between the low- and high-field superconducting phases. 
It is expected that the vortex core size suffers a discontinuity through the transition, which could lead to irreversible effects in the thermodynamics \cite{Higashi2016,mockli2018,schertenleib2021unusual}.  
Another idea would be to examine whether the phase winding of the vortices adopt the registry of the odd-phase. This would imply sublattice dependent $\pi$-shifts in the vortex phases, leading to a new type of twisted phase structure of the Abrikosov flux lines. We speculate that this would lead to new and interesting effects. 

Even though Fig. \ref{fig:full} gives a good fit to the experimental phase diagram, indicating that our treatment is  likely to contain the essential physics, a more complete treatment can also include the pairing mechanism, a more detailed band structure, and the angular dependence of the $g$-factor and Maki parameters. 
We speculate that inclusion of the orbital effect would require a larger $\alpha/t$ ratio for an equivalent fitting. 

Still, from a theoretical perspective, it would be of interest to establish the nature of the pairing mechanism.  In this context, pairing from multipolar Kondo interactions provides an interesting scenario \cite{Patri:2021,Santini:2009,Kuramoto:2009}, given the hidden order observed above the superconducting transition temperature \cite{khim2021fieldinduced}.

From the experimental side, it would be also useful to have more detailed experiments with cleaner samples and lower temperatures in order to determine the presence and location of nodes in the superconducting gap in both low- and high-field phases. Also, magnetization data, which can tell us about properties of the vortex lattice could highlight unusual aspects associated with the high-field phase. Experiments under pressure or strain are also potentially interesting, since these can change the ratios $\alpha/t$ and $\lambda/t$ which should have a clear influence on the critical temperature and upper critical fields that could be traced back within our framework.

\section{Conclusion}

Layered LNCSs are good candidates to observe magnetic field induced two phase superconductivity. CeRh$_2$As$_2$ is a prototypical example. 
From all possible low and high field superconducting phases, we pinpoint a dominant singlet even- to odd transition for CeRh$_2$As$_2$.
The even and odd phases have a rich superconducting wavefunction with both singlet and triplet parts. 
In contrast to centrosymmetric crystals without local noncentrosymmetricity,
singlet-triplet mixing in LNCSs is possible due to
the additional sublattice degree of freedom.
In addition to the usual even-singlet and odd-triplet states, even-triplets in the low-field phase and odd-singlets in the high-field phase can be realized. 
Up to now, phenomenological properties of new superconductors relied on whether the crystal is centrosymmetric or noncentrosymmetric. 
Local noncentrosymmetricity might also lead to new phenomenology under a magnetic field. 

One hallmark of LNCSs  that we have identified in this work is the distinct response of the even and odd phases to impurities. They enhance the critical field of the even phase beyond the Pauli limit, but suppress the odd-phase, leading to a change in the position of the multicritical point in the phase diagram. 
It would be interesting to see if the change in position of the multicritical point can be assessed experimentally. 

Our results provide a suitable starting point for further studies planning to go inside the phase diagram and investigate the first-order phase transition in more detail.
The framework can be readily translated to van der Waals superconductors such as few layer transition metal dichalcogenide and twisted graphene systems.

\begin{acknowledgments}
The authors thank Menashe Haim, Elena Hassinger, Maxim Khodas, Javier Landaeta, Sérgio Magalhães and Andrzej Ptok for enlightening discussions. 
A. R. acknowledges support from the SNSF Ambizione grant. A. R. and D. M.  thank the support of the Condensed Matter Theory Group at the Paul Scherrer Institute that allowed this collaboration to be established.
\end{acknowledgments}

\onecolumngrid
\appendix

\section{Detailed derivations for the even-phase \label{app:low}}

Solving the linearized Eilenberger equations \eqref{eq:linEilenberger} for $\mathbf{B}=(0,0,B)$ and $t_1(\mathbf{k})=t\,\hat{t}_1(\mathbf{k})$ in the clean case, and using the properties of the order parameters listed in Fig. \ref{fig:tabela}, the solution for $f_{31(32)}$ gives
\begin{align}
 f_{31(32)} = \frac{|\omega_n|\left[\pm iB\gamma_{y(x)} d_{00}+\left(\omega_n^2+\gamma_{x(y)}^2+B^2\right)d_{31(32)}+\gamma_x\gamma_yd_{32(31)}\right]}{\left(\omega_n^2+t^2 \right )\left(\omega_n^2+B^2 \right )+\omega_n^2\boldsymbol{\gamma}^2}+(\mbox{terms that vanish in the next average}).
\end{align}
The signs $+(-)$ apply for $f_{31(32)}$. We now use $d_{31(32)}(\mathbf{k})=\eta_{31(32)}\hat{d}_{31(32)}(\mathbf{k})$, where $\hat{d}_{31(32)}(\mathbf{k})=\mp \sqrt{c}\hat{\gamma}_{y(x)}(\mathbf{k})$, with $c=2$, due to the fact that $\{\eta_{31},\eta_{32}\}$ is a pair. 
We retain $c$ to follow the parts that are affected by it.
Applying the same approximation scheme mentioned in Sec. \ref{sec:low}, we obtain the averages
\begin{align}
\langle f_{00} \rangle_\mathbf{k} & =|\omega_n|
\frac{\left(\omega_n^2+\alpha^2+t^2 \right )\eta_{00}+i\sqrt{c}B\alpha\eta_{31(32)}}{(\omega_n^2+t^2)(\omega_n^2+B^2)+\omega_n^2\alpha^2};
\label{eq:00} \\
\left\langle \hat{d}_{31(32)}f_{31(32)}\right\rangle_\mathbf{k} &= 
\frac{|\omega_n|
\left[
 -i\sqrt{c}B\alpha\eta_{00}/2\pm c\alpha^2\langle \hat{\gamma}_x^2\hat{\gamma}_y^2\rangle_\mathbf{k}\left(\eta_{31}-\eta_{32}\right)+c(\omega_n^2+B^2)\eta_{31(32)}/2
 \right]
 }
 {\left(\omega_n^2+t^2 \right )\left(\omega_n^2+B^2 \right )+\omega_n^2\alpha^2}. 
 \label{eq:hd31}
\end{align}
Using $\eta_{31}=\eta_{32}$, the term with $\langle \hat{\gamma}_x^2\hat{\gamma}_y^2\rangle_\mathbf{k}$ does not contribute. 
Let us rewrite Eqs. \eqref{eq:00} and \eqref{eq:hd31} more neatly as
\begin{align}
    \langle f_{00}\rangle_\mathbf{k}=A_1\eta_{00}+i\sqrt{c}A_2\eta_{31(32)};\quad \left\langle \hat{d}_{31(32)}f_{31(32)}\right\rangle_\mathbf{k} = -i\sqrt{c}\frac{A_2}{2}\eta_{00}+cA_3\eta_{31(32)},\label{eq:as}
\end{align}
where $A_1$, $A_2$ and $A_3$ can be identified by comparison with Eqs. \eqref{eq:00} and \eqref{eq:hd31}. 
With no loss of generality,
we choose $\eta_{00}$ to be real and write $\eta_{31(32)}=\mathsf{Re}\eta_{31(32)}+i\mathsf{Im}\eta_{31(32)}$.
Substituting Eqs. \eqref{eq:as} and into their respective self-consistency conditions $\eqref{eq:self_consistency}$, gives
\begin{align}
& \left( \ln\frac{T}{T_{00}}+\mathsf{S}_{00}\right )\eta_{00}+\mathsf{S}_{00,31(32)}\mathsf{Im}\eta_{31(32)}-i\mathsf{S}_{00,31(32)}\mathsf{Re}\eta_{31(32)}=0 \\
& \mathsf{S}_{00,31(32)}\eta_{00}+2\left( \ln\frac{T}{T_{31}}+\mathsf{S}_{31}\right )\mathsf{Im}\eta_{31(32)}-2i\left( \ln\frac{T}{T_{31}}+\mathsf{S}_{31}\right )\mathsf{Re}\eta_{31(32)}=0,
\end{align}
where
\begin{align}
\mathsf{S}_{00}=\pi T\sum_n\left[\frac{1}{|\omega_n|}-A_1 \right ]; \quad 
\mathsf{S}_{00,31(32)}=\pi T\sum_n\sqrt{c}A_2;\quad
\mathsf{S}_{31}=\ln\frac{T}{T_{31}}+\pi T\sum_n\left[\frac{1}{|\omega_n|}-cA_3 \right ].
\end{align}
The Matsubara sums yield Eqs. \eqref{eq:peta} in the main text.
Organizing the coupled equations in matrix form, we obtain
\begin{align}
\begin{bmatrix}
\ln\frac{T}{T_{00}}+\mathsf{S}_{00} & \mathsf{S}_{00,31(32)} & -i\mathsf{S}_{00,31(32)}\\ 
\mathsf{S}_{00,31(32)} & 2\left(\ln\frac{T}{T_{31}}+\mathsf{S}_{31} \right ) & 0\\ 
 0 & 0 & \ln\frac{T}{T_{31}}+\mathsf{S}_{31}
\end{bmatrix}
\begin{bmatrix}
\eta_{00}\\ 
\mathsf{Im}\eta_{31(32)}\\ 
\mathsf{Re}\eta_{31(32)}
\end{bmatrix}=0.
\label{eq:33}
\end{align}

From the determinant of the matrix in Eq. \eqref{eq:33}, one sees that $\eta_{00}$ couples only to $\mathsf{Im}\eta_{31(32)}$. 
One might wonder, what happens to $\mathsf{Re}\,\eta_{31(32)}$? The instability condition is
\begin{align}
\overbrace{\det(\mathsf{P}_\mathsf{even})}^{\mbox{\scriptsize{imaginary part}}}\overbrace{\left(\ln\frac{T}{T_{31}}+\mathsf{S}_{31}\right)}^{\mbox{\scriptsize{real part}}}=0,
\end{align}
where the first term contains $\mathsf{Im}\,\eta_{31(32)}$ and the second term describes $\mathsf{Re}\,\eta_{31(32)}$. Therefore, even if $T_{31}\ll T_{00}$, $\mathsf{Im}\,\eta_{31(2)}$ couples to $\eta_{00}$ (and condenses at $T_{00}$), whereas $\mathsf{Re}\,\eta_{31(2)}$ only condenses at $T_{31}\ll T_{00}$.

\section{Detailed derivations for the odd-phase \label{app:high}}

\subsection{Solutions of the averages}

We look at the same solution set obtained for all $f_{ab}$'s in Appendix \ref{app:low}, but now analyze the  relevant quasiclassical Green's functions for the odd-phase $\{\eta_{30},\eta_{01(02)},\eta_{23}\}$, namely $\{f_{30},f_{01(02)},f_{23}\}$. The solutions are
\begin{align}
f_{30} & =\frac{\left(\omega_n^2+\boldsymbol{\gamma}^2\right)\left(\omega_n^2+\boldsymbol{\gamma}^2+t^2+B^2 \right )d_{30}
+iB\left(\omega_n^2+\boldsymbol{\gamma}^2+B^2-t^2 \right )\left(\gamma_x d_{02}-\gamma_y d_{01} \right )
+2iBt\left(\omega_n^2+\boldsymbol{\gamma}^2 \right )d_{23}}
{|\omega_n|\left[\omega_n^2+\boldsymbol{\gamma}^2+(B+t)^2 \right ]\left[\omega_n^2+\boldsymbol{\gamma}^2+(B-t)^2 \right ]}; 
\end{align}
\small
\begin{align}
f_{01(02)}  &=\frac{
\pm iB\gamma_{y(x)}\left(\omega_n^2+\boldsymbol{\gamma}^2+B^2-t^2 \right )d_{30}+\left[ \left(\omega_n^2+\gamma_{x(y)}^2+t^2 \right )\left(\omega_n^2+\boldsymbol{\gamma}^2+t^2\right)+B^2\left(2\omega_n^2+2\gamma_{x(y)}^2+\gamma_{y(x)}^2-2t^2 \right )+B^4\right ]d_{01(02)}}
{|\omega_n|\left[\omega_n^2+\boldsymbol{\gamma}^2+(B+t)^2 \right ]\left[\omega_n^2+\boldsymbol{\gamma}^2+(B-t)^2 \right ]} \notag \\
& + \frac{
\gamma_x\gamma_y\left(\omega_n^2+\boldsymbol{\gamma}^2+B^2+t^2 \right )d_{02(01)}
\pm t\gamma_{y(x)}\left(\omega_n^2+\boldsymbol{\gamma}^2+t^2-B^2 \right )d_{23}}
{|\omega_n|\left[\omega_n^2+\boldsymbol{\gamma}^2+(B+t)^2 \right ]\left[\omega_n^2+\boldsymbol{\gamma}^2+(B-t)^2 \right ]};
\end{align}
\normalsize
\begin{align}
f_{23}=\frac{-2itB\left(\omega_n^2+\boldsymbol{\gamma}^2 \right )d_{30}
+t\left(\omega_n^2+\boldsymbol{\gamma}^2+t^2-B^2 \right )\left(\gamma_yd_{01}-\gamma_xd_{02} \right )+\left(\omega_n^2+\boldsymbol{\gamma}^2\right)\left(\omega_n^2+\boldsymbol{\gamma}^2+t^2+B^2 \right )d_{23}}
{|\omega_n|\left[\omega_n^2+\boldsymbol{\gamma}^2+(B+t)^2 \right ]\left[\omega_n^2+\boldsymbol{\gamma}^2+(B-t)^2 \right ]}.
\end{align}
We omitted additional terms in the solutions that later vanish in the averages.
Looking at these solutions, one can readily identify the parameters that couple different order parameters according to the wheel in Fig. \ref{fig:wheel}.
We now calculate the relevant averages that enter the self-consistency condition Eq. \eqref{eq:self_consistency}. To avoid the exhaustive repetition of the denominator, let us define
\begin{align}
\mathsf{D}= |\omega_n|\left[\omega_n^2+\alpha^2+\left(B+t\right)^2\right]\left[\omega_n^2+\alpha^2+\left(B-t\right)^2\right].
\end{align}
We now write $d_{30}(\mathbf{k})=\eta_{30}$, $d_{01(02)}(\mathbf{k})=\eta_{01(02)}\hat{d}_{01(02)}(\mathbf{k})$ with $\hat{d}_{01(02)}(\mathbf{k})=\mp \sqrt{c_\gamma}\hat{\gamma}_{y(x)}(\mathbf{k})$, and
$d_{23}(\mathbf{k})=\eta_{23}\hat{d}_{23}(\mathbf{k})$ with $\hat{d}_{23}(\mathbf{k})=\sqrt{c_t}\hat{t}_1(\mathbf{k})$. Here $c_\gamma=2$ and $c_t=1$. 
The averages that enter the self-consistency condition are 
\begin{align}
\mathsf{D}\langle f_{30}\rangle  =\left(\omega_n^2+\alpha^2\right)\left(\omega_n^2+\alpha^2+B^2+t^2\right)\eta_{30}+iB\left[\frac{\sqrt{c_\gamma}}{2}\alpha\left(\omega_n^2+\alpha^2+B^2-t^2\right)\left(\eta_{01}+\eta_{02} \right )+ 2\sqrt{c_t}t\left(\omega_n^2+\alpha^2\right)\eta_{23}\right]; \label{eq:c30}
\end{align}
\small
\begin{align}
\mathsf{D}\langle\hat{d}_{01(02)}f_{01(02)}\rangle  & = -\frac{i\sqrt{c_\gamma}B\alpha}{2}\left(\omega_n^2+\alpha^2+B^2-t^2\right)\eta_{30}
+\frac{c_\gamma}{2}\left[\left(\omega_n^2+\frac{\alpha^2}{2}+t^2\right)\left(\omega_n^2+\alpha^2+t^2\right)+2B^2\left(\omega_n^2+\frac{3\alpha^2}{4}-t^2\right)+B^4\right]\eta_{01(02)} \notag \\
& -c_\gamma \alpha^2\left(\omega_n^2+\alpha^2+B^2+t^2 \right )\langle \hat{\gamma}_x^2\hat{\gamma}_y^2\rangle\eta_{02(01)} 
-\sqrt{c_\gamma c_t}\alpha t \langle\hat{t}_1^2\hat{\gamma}_{y(x)}^2\rangle\left(\omega_n^2+\alpha^2+t^2-B^2 \right )\eta_{23};
\label{eq:c01}
\end{align}
\normalsize
\begin{align}
\mathsf{D}\langle \hat{d}_{23}f_{23}\rangle & =-2i\sqrt{c_t}Bt(\omega_n^2+\alpha^2)\eta_{30}
-\sqrt{c_tc_\gamma}\alpha t\left(\omega_n^2+\alpha^2+t^2-B^2\right)\left(\langle \hat{t}_1\hat{\gamma}_y^2\rangle\eta_{01}+\langle \hat{t}_1\hat{\gamma}_x^2\rangle\eta_{02} \right ) \notag \\
&+c_t(\omega_n^2+\alpha^2)\left(\omega_n^2+\alpha^2+B^2+t_1^2\right)\eta_{23}.
\label{eq:c23}
\end{align}
The averages involving two square terms evaluate to $\langle \hat{\gamma}_x^2\hat{\gamma}_y^2\rangle_\mathbf{k}=1/16$ and $\langle\hat{t}_1^2\hat{\gamma}_{y(x)}^2\rangle_\mathbf{k}=1/4$. Let us rewrite the averages more neatly as

\begin{align}
\langle f_{30}\rangle &  =B_1\eta_{30}+i\sqrt{c_\gamma}B_2\left(\eta_{01}+\eta_{02} \right )+i\sqrt{c_t}B_3\eta_{23}; \\
\langle \hat{d}_{01(02)}f_{01(02)}\rangle & =-i\sqrt{c_\gamma}B_2\eta_{30}+c_\gamma B_4\eta_{01(02)}-c_\gamma B_5\eta_{02(01)}-\sqrt{c_\gamma c_t}B_6 \eta_{23}; \\
\langle \hat{d}_{23}f_{23}\rangle & =-i\sqrt{c_t}B_3\eta_{30}-\sqrt{c_\gamma c_t}B_6\left(\eta_{01}+\eta_{02} \right )+c_tB_1\eta_{23}.
\end{align}
The $B_i$'s ($i=1\ldots 6$) can be identified by comparison with Eqs. \eqref{eq:c30}, \eqref{eq:c01} and \eqref{eq:c23}. The six $B_i$'s lead to six distinct Matsubara sums.

\subsection{Matsubara sums}

Writing the real and imaginary parts relative to $\eta_{30}$, and using the self-consistency condition Eq. \eqref{eq:self_consistency}, we identify the following Matsubara sums:
\begin{align}
\mathsf{S}_{30} & =\pi T\sum_{n\in\mathbb{Z}}\left[\frac{1}{|\omega_n|}-B_1\right] =\pi T\sum_{n\in\mathbb{Z}}\left[\frac{1}{|\omega_n|}-\frac{\left(\omega_n^2+\alpha^2\right)\left(\omega_n^2+\alpha^2+B^2+t^2\right)}{|\omega_n|\left[\omega_n^2+\alpha^2+\left(B+t\right)^2\right]\left[\omega_n^2+\alpha^2+\left(B-t\right)^2\right]}\right];\label{eq:SA} \\
 \mathsf{S}_{01(02)} & 
= \pi T\sum_{n\in\mathbb{Z}}c_\gamma B_4= \pi T\sum_{n\in\mathbb{Z}}\left[\frac{1}{|\omega_n|}-\frac{c_\gamma}{2}\frac{\left(\omega_n^2+\frac{\alpha^2}{2}+t^2\right)\left(\omega_n^2+\alpha^2+t^2\right)+2B^2\left(\omega_n^2+\frac{3\alpha^2}{4}-t^2\right)+B^4}{|\omega_n|\left[\omega_n^2+\alpha^2+\left(B+t\right)^2\right]\left[\omega_n^2+\alpha^2+\left(B-t\right)^2\right]}\right]; \label{eq:convergence} \\
\mathsf{S}_{01,02} & = \pi T\sum_{n\in\mathbb{Z}}c_\gamma B_5= \pi T\sum_{n\in\mathbb{Z}}\frac{c_\gamma}{16}\frac{\alpha^2\left(\omega_n^2+\alpha^2+B^2+t^2 \right )}{|\omega_n|\left[\omega_n^2+\alpha^2+\left(B+t\right)^2\right]\left[\omega_n^2+\alpha^2+\left(B-t\right)^2\right]}\overset{c_\gamma=2}{=}\frac{\mathsf{S}_{01(02)}}{4}; \\
\mathsf{S}_{30,01(02)} &=\pi T\sum_{n\in\mathbb{Z}}\sqrt{c_\gamma}B_2= 
\frac{\pi T}{2}\sum_{n\in\mathbb{Z}}\frac{\sqrt{c_\gamma}B\alpha \left(\omega_n^2+\alpha^2+B^2-t^2\right) }{|\omega_n|\left[\omega_n^2+\alpha^2+\left(B+t\right)^2\right]\left[\omega_n^2+\alpha^2+\left(B-t\right)^2\right]}; \label{eq:s3001}\\
\mathsf{S}_{30,23} & 
=\pi T\sum_{n\in\mathbb{Z}}\sqrt{c_t}B_3=\pi T\sum_{n\in\mathbb{Z}} \frac{2\sqrt{c_t}B t\left(\omega_n^2+\alpha^2\right)}{|\omega_n|\left[\omega_n^2+\alpha^2+\left(B+t\right)^2\right]\left[\omega_n^2+\alpha^2+\left(B-t\right)^2\right]}; \\
\mathsf{S}_{01(02),23} & 
=\pi T\sum_{n\in\mathbb{Z}}\sqrt{c_\gamma c_t}B_6=\pi T\sum_{n\in\mathbb{Z}} \frac{\sqrt{c_\gamma c_t}}{4}\frac{\alpha t
\left(\omega_n^2+\alpha^2-B^2+t^2\right)}{|\omega_n|\left[\omega_n^2+\alpha^2+\left(B+t\right)^2\right]\left[\omega_n^2+\alpha^2+\left(B-t\right)^2\right]}. \label{eq:A1uA2u}
\end{align}
The Matsubara sum in Eq. \eqref{eq:convergence} is convergent for $c_\gamma=2$. 
In contrast to the Matsubara sums of the even-phase, all Matsubara sums of the odd-phase can be performed analytically without appealing to root functions.
They are listed below:
\begin{align}
\mathsf{S}_{30} = \frac{1}{2}\sum_{l=\pm}\frac{\left(B+l t \right )^2}{\alpha^2+\left(B+lt \right )^2}\,\mathsf{Re}\, H\left(-\frac{1}{2}+i\frac{\sqrt{\alpha^2+\left(B+lt \right )^2}}{2\pi T} \right )
+\frac{\alpha^2\left(B^2+t^2 \right )+\left(B^2-t^2 \right )^2}{\left[\alpha^2+\left(B+t\right )^2 \right ]\left[\alpha^2+\left(B-t \right )^2 \right ]}\ln 4.
\label{eq:sa}
\end{align}

\begin{align}
\mathsf{S}_{01(02)} = \frac{1}{4}\sum_{l=\pm}\frac{\alpha^2}{\alpha^2+\left(B+lt \right )^2}\mathsf{Re}\, H\left(-\frac{1}{2}+i\frac{\sqrt{\alpha^2+\left(B+lt \right )^2}}{2\pi T} \right )
+\frac{\alpha^2\left(\alpha^2+B^2+t^2 \right )\ln 2}{\left[\alpha^2+\left(B+t \right )^2 \right ]\left[\alpha^2+\left(B-t \right )^2 \right ]}.
\end{align}
\begin{align}
    \mathsf{S}_{30,01(02)} =\frac{1}{4\sqrt{2}} \sum_{l=\pm}\frac{\alpha\left(B+lt \right )}{\alpha^2+\left(B+lt \right )^2}\,\mathsf{Re}\, H\left(-\frac{1}{2}+i\frac{\sqrt{\alpha^2+\left(B+lt \right )^2}}{2\pi T} \right )
+
\frac{1}{\sqrt{2}}\frac{\alpha B\left(\alpha^2+B^2-t^2 \right )}{\left[\alpha^2+\left(B+t \right )^2 \right ]\left[\alpha^2+\left(B-t \right )^2 \right ]}.
\end{align}
\begin{align}
\mathsf{S}_{30,23} = \frac{1}{2}\sum_{l=\pm}\frac{l\left(B+lt \right )^2}{\alpha^2+\left(B+lt \right )^2}\,\mathsf{Re}\, H\left(-\frac{1}{2}+i\frac{\sqrt{\alpha^2+\left(B+lt \right )^2}}{2\pi T} \right )
+
\frac{B t\alpha^2\ln 16}{\left[\alpha^2+\left(B+t \right )^2 \right ]\left[\alpha^2+\left(B-t \right )^2 \right ]}.
\end{align}
\begin{align}
\mathsf{S}_{01(02),23} = \frac{1}{4\sqrt{2}}\sum_{l=\pm}\frac{l\left(B+lt \right )^2}{\alpha^2+\left(B+lt \right )^2}\,\mathsf{Re}\, H\left(-\frac{1}{2}+i\frac{\sqrt{\alpha^2+\left(B+lt \right )^2}}{2\pi T} \right )
+\frac{1}{\sqrt{2}}
\frac{\alpha t\left(\alpha^2-B^2+t^2 \right )\ln 2}{\left[\alpha^2+\left(B+t \right )^2 \right ]\left[\alpha^2+\left(B-t \right )^2 \right ]}.
\label{eq:seg}
\end{align}
Here, $H(z)=\sum_{k=1}^zk^{-1}$ is the Harmonic number, which is related to the digamma function $\psi(z)$ via 
\begin{align}
H\left(-\frac{1}{2}+z\right)=\psi\left(\frac{1}{2}+z\right)+\gamma,
\end{align}
where $\gamma$ is the Euler-Mascheroni constant. One can develop $\alpha\gg t$ and $t\gg \alpha$ expansions if one is interested in pushing analytic results further. An expansion in $B$ might also be useful to derive a Ginzburg-Landau theory.

\subsection{The instability condition}

Rewriting the self-consistency condition in matrix form and identifying the Matsubara sums, we obtain
\begin{align}
\begin{bmatrix}
\ln\frac{T}{T_{30}}+\mathsf{S}_{30} & \mathsf{S}_{30,01} & \mathsf{S}_{30,02} & \mathsf{S}_{30,23} & -i\mathsf{S}_{30,01} & -i\mathsf{S}_{30,02} & -i\mathsf{S}_{30,23}\\ 
\mathsf{S}_{30,01} & \ln\frac{T}{T_{01}}+\mathsf{S}_{01} & \mathsf{S}_{01,02} & \mathsf{S}_{01,23} 
 & 0 & 0 & 0\\ 
\mathsf{S}_{30,02} & \mathsf{S}_{01,02} & \ln\frac{T}{T_{02}}+\mathsf{S}_{02} & \mathsf{S}_{02,23} & 0 & 0 & 0\\ 
\mathsf{S}_{30,23} & \mathsf{S}_{01,23} & \mathsf{S}_{02,23} & \ln\frac{T}{T_{23}}+\mathsf{S}_{30} & 0 & 0 & 0\\ 
 0 & 0 & 0 & 0 & \ln\frac{T}{T_{01}}+\mathsf{S}_{01} & \mathsf{S}_{01,02} & \mathsf{S}_{01,23}\\
0 & 0 & 0 & 0 & \mathsf{S}_{01,02} & \ln\frac{T}{T_{02}}+\mathsf{S}_{02} & \mathsf{S}_{02,23}\\ 
0 & 0 & 0 & 0 & \mathsf{S}_{01,23} & \mathsf{S}_{02,23} & \ln\frac{T}{T_{23}}+\mathsf{S}_{30} 
\end{bmatrix}
\begin{bmatrix}
\eta_{30}\\ 
\mathsf{Im}\eta_{01}\\ 
\mathsf{Im}\eta_{02}\\ 
\mathsf{Im}\eta_{23}\\ 
\mathsf{Re}\eta_{01}\\ 
\mathsf{Re}\eta_{02}\\ 
\mathsf{Re}\eta_{23}
\end{bmatrix}=0.
\end{align}
The pair-breaking equation is given by the determinant of the matrix above, which is given in Eq. \eqref{eq:pair_breakinh_high}. The magnetic field only couples the imaginary parts of the triplet components to $\eta_{30}$.

\twocolumngrid
\newpage 

\bibliography{bibliography}

\end{document}